\documentclass[nofigure]{pasj01}
\Received{$\langle$02-Jun-2019$\rangle$}
\Accepted{$\langle$25-Jan-2020$\rangle$}
\Published{$\langle$publication date$\rangle$}

\usepackage{times}
\usepackage[pdftex]{graphicx,color}

\newcounter{num}
\newcommand{\Rnum}[1]{\setcounter{num}{#1} \Roman{num}}

\begin{document}

\title{Polarization imaging of \textcolor{black}{M87} jets by general relativistic radiative transfer calculation based on GRMHD simulations}

\author{Yuh \textsc{Tsunetoe}\altaffilmark{1}$^*$, Shin \textsc{Mineshige}\altaffilmark{1}, Ken \textsc{Ohsuga}\altaffilmark{2}, Tomohisa \textsc{Kawashima}\altaffilmark{3}, Kazunori \textsc{Akiyama}\altaffilmark{4,5,3,6}}

\altaffiltext{1}{Department of astronomy, Kyoto University, Kitashirakawa-Oiwake-cho, Sakyo-ku, Kyoto, 606-8502}
\altaffiltext{2}{Center for Computational Sciences, University of Tsukuba, 1-1-1 Tennodai, Tsukuba, Ibaraki, 305-8577}
\altaffiltext{3}{National Astronomical Observatory of Japan, 2-21-1 Osawa, Mitaka-shi, Tokyo,  181-8588}
\altaffiltext{4}{National Radio Astronomy Observatory, 520 Edgemont Rd, Charlottesville, VA, 22903, USA}
\altaffiltext{5}{Massachusetts Institute of Technology, Haystack Observatory, 99 Millstone Road, Westford, MA 01886, USA}
\altaffiltext{6}{Black Hole Initiative, Harvard University, 20 Garden Street, Cambridge, MA 02138, USA}

\email{tsunetoe@kusastro.kyoto-u.ac.jp}

\maketitle

\begin{abstract}

The spectacular images of the M87 black hole taken by the Event Horizon Telescope (EHT) have opened a new era of black hole research. One of the next issues is to take polarization images around the central black hole (BH). Since radio emission is produced by synchrotron process,
polarization properties should vividly reflect the magnetic field structures at the jet base and thus provide good information regarding the magnetic mechanism of jet formation.
With this kept in mind we perform general relativistic (GR) radiative transfer calculations of polarized light based on the GR magnetohydrodynamic (MHD) simulation data of accretion flow and outflow \textcolor{black}{in M87}, to obtain their linear and circular polarization images in the BH horizon-scale.
We found that the linear polarization components originating from the jet base and inner accretion flow should experience Faraday rotation and depolarization when passing through magnetized plasmas around the BH, thus sensitively depending on the BH spin.
Through the comparison with \textcolor{black}{total intensity image at 1.3~mm by EHT and the polarization degree and the rotation measure (RM) measured at 1.3~mm with the Submillimeter Array}, the model with the spin parameter of $a=0.9M_\mathrm{BH}$ (with $M_\mathrm{BH}$ being the BH mass) is favored over other models with $a = 0.5 M_\mathrm{BH}$ or $0.99 M_\mathrm{BH}$, though we need further systematic studies for confirmation.
\textcolor{black}{We also find in low-temperature models clear ring-like image in the circular polarization map, which arises because of Faraday conversion of the linearly polarized synchrotron emission \textcolor{black}{and is thus indicative of magnetic field direction}. This occurs only when the emission region is threaded with well-ordered magnetic fields and hence no clear images are expected in high-temperature disk models, in which disk emission is appreciable.
}
We will be able to elucidate the field configuration through the comparison between the simulated polarization images and future polarimetry with EHT and other VLBI observations.

\end{abstract}

\section{Introduction}

It had been a widely accepted hypothesis that 
accretion onto a supermassive black hole (SMBH) produces 
enormous power from active galactic nuclei (AGNs).
Recently, this hypothesis has been directly proven 
by observations with the Event Horizon Telescope (EHT),
a global VLBI (very long baseline interferometry) array
with an unprecedented spatial resolution of $\sim25$ $\mu$as (although other possibilities are not completely ruled out, \cite{EHT19a}).
The EHT has captured the first horizon-scale images of the central SMBH in the low luminosity AGN (LLAGN; \cite{Nag00}) M87 at 230~GHz (\cite{EHT19a}).
The first obtained images 
exhibit a shadow of the black hole (\cite{Hi16}; \cite{von21}) surrounded by a photon ring (\cite{Bar73}; \cite{Fa00}; \cite{Ta04}), for the BH silhouette with a luminous accretion disk, see \citet{Lu79} and \citet{FY88}.

As is well known, AGNs are the persistent origin of high energy phenomena, 
such as intense and variable radiation in all wavelengths and powerful outflow, 
thereby affecting the evolution of their host galaxies and member stars. 
In addition, AGNs are occasionally associated with relativistic, collimated jets (\cite{BK79}),
though the acceleration and collimation mechanisms of the jets still remain open questions. 
There are plenty of ideas that have been proposed;
e.g., relativistic jets could be driven by the extraction of black hole spin energy (\cite{BZ77}),
but the relationship between the presence of relativistic jets and the black hole spin 
is still controversial.
This is definitely one of the most outstanding unresolved issues in modern astronomy. 
It is interesting and also surprising, in a sense, to note in this context 
that no clear jet images were obtained in the EHT observation of M87, 
despite the fact that M87 is famous in exhibiting extended jet features 
in optical and in longer wavelength radio bands.
Sparse coverages of the 2017 EHT array do not provide high dynamic ranges for the images enough to capture the more extended jet emission, which becomes fainter and optically thinner than longer observing wavelengths.

The most promising models to date for producing AGN jets are those making use of 
magnetic fields. Hence, an important clue to resolving the jet enigma should be provided 
by magnetic-field structure at the launching point of the jet close to the SMBH. 
Magnetic fields are also known to play a predominant role in accretion flow
and for this reason multi-dimensional MHD (magnetohydrodynamical) simulations 
have been performed rather extensively in these days by many groups (e.g. \cite{Mo16}; \cite{Ch19}).
We should keep in mind that such theoretical models of accretion flows and relativistic jet have uncertainties (such as initial magnetic field configurations) 
and that the results may sensitively depend on model parameters 
(e.g., prescriptions to calculate electron temperatures; \cite{EHT19e}).
We should point that observational appearance of accretion flow and jets may drastically 
vary, as a viewing angle changes.
But nevertheless some GRMHD (general relativistic magneto-hydrodynamic) simulations
could successfully reproduce the main observational features of the M87 jet (\cite{Na18}; \cite{Ch19}; \cite{EHT19e}). 
In order to mediate between such models and observations, 
it will very useful to calculate theoretical images of AGN accretion flow and outflow
by solving radiative transfer problems (\cite{De12}; \cite{Mo16}; \cite{Ch19}).

In the present study, we focus on the polarization properties of the black hole images.
It is known that polarization images calculated by polarized radiative transfer vividly reflect magnetic structure in plasma (\cite{BL09}; \cite{De16}; \cite{Ch16}; \cite{Mo17}; \cite{An18}).
\textcolor{black}{There are some previous studies of polarization in the core region of AGNs, \citet{Mo17} focused on the Faraday rotation in M87 jet. In their SANE (standard and normal evolution) models, emission in the counter-jet (i.e., the jet receding from the observer) is dominant for 230 GHz image, and depolarized when passing through the accretion disk. As a result, the foreground (approaching) jet significantly contributes to linear polarization.}
Comparing such theoretical images with those obtained by the VLBI observations with superb spatial resolution, 
we can elucidate actual magnetic structure around the SMBH.
In fact, we could see the base of jet near to the black hole 
in 22, 43 and 86~GHz VLBI observations, 
and the horizon-scale disk and/or jet with the EHT at 230, 345 and 690~GHz. 
From these observation, we will be able to verify 
the Blandford-Znajek process (\cite{BZ77}; \cite{Be09}; \cite{Tch11}), which was strongly favored by parameter analysis in the EHT observation (\cite{EHT19e}), and/or the Blandford-Payne process (\cite{BP82}), 
promising mechanisms of creating relativistic jets and/or outflow in AGN.

The typical scenario for the origin of polarized radio emission in AGN is as follows:
first, linearly polarized radiation is produced by synchrotron emission 
in a compact, energetic region; i.e., in the inner disk and/or at the jet base.
Such linearly polarized light then experiences Faraday effects, 
especially Faraday rotation of linear polarization (e.g. \cite{Ag00}; \cite{Ma06}; \cite{Kuo14}),
when it passes through outer, magnetized regions, so-called the Faraday screen. 
As a result, we expect that linearly polarized light should be depolarized significantly,
and that the initial information, e.g., magnetic field configuration, is lost. 
However, the actual situation may not be so simple. 
Synchrotron emission arises in the Faraday screen, and the Faraday effects 
manifest themselves even in such energetic region as the jet base. 
Therefore, we need to consider all of these polarization process in the whole region 
for precisely calculating polarization properties to be compared with polarimetry data.

With such thoughts kept in mind, we performed general relativistic radiative transfer (GRRT) simulations, taking into account full polarization in mm-submm wavelengths. 
The radiation in these wavelength ranges is characterized by high polarization degrees 
owing to substantial synchrotron emission from relativistic plasma near to the black hole. 
Since the polarization properties can be described in terms of the four Stokes parameters, 
we here simulate synchrotron emission, synchrotron self-absorption of each polarization component, and the Faraday effects between Stokes parameters (i.e., the Faraday rotation and the Faraday conversion). 
The polarization image obtained here reflects the distorted black hole spacetime, 
as well as the magnetic field configurations and the bulk motions of the plasma.
We thus expect to find the black-hole spin dependence of the observed polarization images.

\textcolor{black}{
There are several differences between \citet{Mo17} and the present paper: First of all, we pay special attention to the properties of the circular polarization, which are amplified by the Faraday conversion of the linear polarization, in addition to the liner polarization. We will demonstrate in the present study that the circular polarization, 
as well as the linear polarization, contains  important clues to resolving 
the magnetic-field  structure in the emission region near the BH. Next, our axi-symmetric semi-MAD jets are stronger than SANE jets and are consistent with multi-wavelengths observations in these shape (\cite{Na18}), and this enables us to compare the horizon-scale image with EHT observation more consistently.
Finally, we also examine the BH spin dependence of the linear polarization maps.}

This paper is organized in the following way:
In section 2, we introduce our models based on the GRMHD simulation data by \citet{Na18} and describe our methods of polarized radiative transfer calculation. 
We then show the results of the calculated polarization images in various parameter ranges in section 3
and present discussion regarding the comparison with observations etc in section 4. 
A final section is devoted to conclusions.

\section{Our Models and Methods of Calculations}
Our study is in two parts: 
(1) GRMHD simulations with data taken from \citet{Na18} and 
(2) polarized GRRT calculations. 
We will describe each of them in the following two sections.

\subsection{GRMHD simulation model of LLAGNs with jet}

\citet{Na18} simulated axisymmetric jet structure, starting with a weakly magnetized torus around a BH, by using the HARM code (\cite{HARM}). 
By adjusting parameters, such as the magnetic configuration of the jets and the minimum plasma-$\beta$ in torus, they succeeded in reproducing
a converging, quasi-stationary jet with a parabolic streamline of $z \propto R^{1.6}$
(with $z$ and $R$ being the half-width of the jet and distance from the black hole), 
in excellent agreement with the VLBI observations (\cite{AN12}; \cite{Do12}; \cite{NA13}; \cite{Ha13}; \cite{Ak15}; \cite{Ha16}, and figure 15 in \cite{Na18}).

\textcolor{black}{The `phi' values of $\phi \equiv \frac{\Phi}{\sqrt{\dot{M}}}$, which is defined as the normalized magnetic flux crossing the BH, are $\sim 30$ in our models (Nakamura et al.~2018), near to but a bit less than a popular definition of MAD state of  40-60. Therefore we refer the models as `semi-MAD'. \textcolor{black}{The magnetic field strength reaches its maximum of $\approx 50$ Gauss near the black hole in our fiducial model. This value is consistent with the values obtained by the previous works regarding the M87 core region; \citet{Ki15} report $50 {\rm G} \le B_{\rm tot} \le 124 {\rm G}$, while \citet{Ch19} report $|B|_{r=10r_{\rm g}} \approx 20 {\rm G}$.}}

The model parameters of the GRMHD simulations are the black hole mass, $M_\mathrm{BH}$, 
and the black hole spin, $a_\mathrm{BH}$.
The mass accretion rate in the horizon $\dot{M}$ is not a free parameter,
since we adjust this so as to reproduce the observed intensity of M87 (explained in \ref{Cmodels}).
Note that density normalization can be taken arbitrarily in non-radiative MHD simulations,
since density and field strength square can be scaled, as long as we fix their ratio.

\subsection{Polarized radiative transfer calculation}

\begin{table*}[t]
\begin{center}
  \begin{tabular}{lccc|c||cc|l}
    Reference name & $a_\mathrm{BH} [M_\mathrm{BH}]$ & $R_\mathrm{high}$ & $i$ & $\dot{M} [M_\odot/\mathrm{yr}]$ & $\pi_{230}$ & $\mathrm{RM}_{\sim230}$ [$\mathrm{rad/m^2}$] & Remarks \\ \hline
    a09R100 & $0.9$ & $100$ & $160\degree$ & $1.4\times10^{-3}$ & $2.3\%$ & $-2.9\times10^5$ & fiducial model \\
    a05R100 & $0.5$ & $100$ & $160\degree$ & $6.5\times10^{-2}$ & $1.1\%$ & $-6.9\times10^6$ & low BH spin model \\
    a099R100 & $0.99$ & $100$ & $160\degree$ & $1.0\times10^{-3}$ & $5.9\%$ & $-4.5\times10^5$ & high BH spin model\\
    a09R10 & $0.9$ & $10$ & $160\degree$ & $9.0\times10^{-4}$ & $1.4\%$ & $-3.2\times10^5$ & hot disk model\\
    a09R100-i135 & $0.9$ & $100$ & $135\degree$ & $1.1\times10^{-3}$ & $0.82\%$ & $1.1\times10^7$ & nearly edge-on model\\  \hline
    
    a09R100-i20 & $0.9$ & $100$ & $20\degree$ & $1.5\times10^{-3}$ & $1.5\%$ & $6.3\times10^4$ & no image \\
    a05R100-i20 & $0.5$ & $100$ & $20\degree$ & $1.1\times10^{-1}$ & $0.55\%$ & $-9.2\times10^6$ & no image \\
    a099R100-i20 & $0.99$ & $100$ & $20\degree$ & $1.0\times10^{-3}$ & $13\%$ & $-2.5\times10^4$ & no image\\
    a09R10-i20 & $0.9$ & $10$ & $20\degree$ & $1.0\times10^{-3}$ & $1.6\%$ & $-1.2\times10^6$ & no image \\
    a09R100-i45  & $0.9$ & $100$ & $45\degree$ & $1.2\times10^{-3}$ & $1.1\%$ & $8.1\times10^6$ & no image \\  \hline
          
  \end{tabular}
\end{center}
  \caption{Calculated models and calculated mass accretion rate, $\dot{M}$, polarization fraction, $\pi=\sqrt{Q^2+U^2+V^2}/I$, and rotation measure (RM) calculated from 230 \& 235~GHz simulations. In all models we fix the black hole mass to be \textcolor{black}{$M_\mathrm{BH}=6.5\times10^9M_\odot$} and temperature ratio between electron-proton in low-$\beta$ region to be $R_\mathrm{low}=1$. Free parameters are the black hole spin $a_\mathrm{BH}$, temperature ratio in high-$\beta$ region $R_\mathrm{high}$ and inclination angle $i$. The mass accretion rate $\dot{M}$ is a scaling parameter to the 230~GHz-observed flux of M87, \textcolor{black}{$\approx 0.5{\rm Jy}$}.}
  \label{models}
\end{table*}

We perform 3D-GR ray-tracing in Boyer-Lindquist coordinates 
$(t,r,\theta,\phi)$ by means of 4th order Runge-Kutta method, rewinding time in infinity 
$t$ from pixels on the observer's screen to plasma region 
according to the following equations (here, we take $c=G=M_\mathrm{BH}=1$);
\begin{equation}
	\Sigma\frac{dt}{d\lambda} = -a(aE\mathrm{sin}^2\theta-L) + \frac{(r^2+a^2)T}{\Delta},
\end{equation}
\begin{equation}
	\Sigma\frac{dr}{d\lambda} = \pm\sqrt{V_r},
\end{equation}
\begin{equation}
	\Sigma\frac{d\theta}{d\lambda} = \pm\sqrt{V_\theta},
\end{equation}
\begin{equation}
	\Sigma\frac{d\phi}{d\lambda} = -(aE-\frac{L}{\mathrm{sin}^2\theta}) + \frac{aT}{\Delta},
\end{equation}
\begin{eqnarray}
	T &=& E(r^2+a^2) - La, \\
	V_r &=& T^2 - \Delta[\mu^2r^2 + (L-aE)^2 +Q], \\
	V_\theta &=& Q - \mathrm{cos}^2\theta\left[a^2(\mu^2-E^2) + \frac{L^2}{\mathrm{sin}^2\theta}\right],
\end{eqnarray}
where $\lambda$ is an affine parameter, $E$, $L$, and $Q$ are the energy, angular momentum, and Carter constant of a particle (\cite{Ca68}), respectively (\cite{Bar72}; \cite{Ch83}; \cite{MM15}). We let $\mu=0$ for photons.
 
Once we find the path of each ray, we can integrate each polarization component
expressed by the Stokes parameters $(I,Q,U,V)$ along the ray from plasma to the screen. 
The equations of polarized radiative transfer are;
\begin{equation}
	\frac{d\mathcal{I}}{d\lambda} = \mathcal{J} - \mathcal{K}\mathcal{I}.
\end{equation}
Here $\lambda$ is common with GR ray-tracing, and
\begin{equation}
	\mathcal{I} = g^3
		\left(
		\begin{array}{c}
			I \\
			Q \\
			U \\
			V \\
		\end{array}
		\right),
\end{equation}
\begin{equation}
	\mathcal{J} = g^2
		\left(
		\begin{array}{c}
			j_I \\
			j_Q \\
			j_U \\
			j_V \\
		\end{array}
		\right),
\end{equation}
\begin{equation}	
	\mathcal{K} = g^{-1}
		\left(
		\begin{array}{cccc}
			\alpha_I & \alpha_Q & \alpha_U & \alpha_V \\
			\alpha_Q & \alpha_I & \rho_V & \rho_U \\
			\alpha_U & -\rho_V & \alpha_I & \rho_Q \\
			\alpha_V & -\rho_U & -\rho_Q & \alpha_I \\
		\end{array}
		\right),
		\label{poleq}
\end{equation}
where $g \equiv \nu_\mathrm{obs}/\nu_\mathrm{em}$,
are invariant forms in plasma rest frame coordinates by tetrad basis 
$(e_{\left(t\right)}^\mu,e_{\left(r\right)}^\mu,e_{\left(\theta\right)}^\mu,e_{\left(\phi\right)}^\mu)$ (\cite{Kro05}; \cite{SH11}; \cite{Ku11}; \cite{De16}). 
We implemented polarized radiative coefficients $(j_*,\alpha_*,\rho_*)$ of 
synchrotron electrons with ultra-relativistic thermal distribution from previous works 
(\cite{Ma96}; \cite{Sh08}; \cite{De16}). 
Finally we obtain the polarization images from integrated $\mathcal{I}$ on the screen in the observer's frame.

Faraday coefficients $(\rho_Q,\rho_U,\rho_V)$ are often so large in the disk region that 
the calculation are practically impossible. To avoid such difficulty, 
we adopted the alternative expression of polarized components $(Q,U,V)$, spherical Stokes parameter $(R_S,\Psi_S,\Phi_S)$ (\cite{Sh12});
\begin{equation}
	Q = R_S\mathrm{sin}\Psi_S\mathrm{cos}\Phi_S,
\end{equation}
\begin{equation}
	U = R_S\mathrm{sin}\Psi_S\mathrm{sin}\Phi_S,
\end{equation}
\begin{equation}
	V = R_S\mathrm{cos}\Psi_S,
\end{equation}
and we can then successfully calculate polarization fraction 
without divergence nor attenuation (\cite{Mo17}).

We set the screen at $r=10^4r_\mathrm{g}$ from the central black hole
and trace rays with the black hole being at the origin. 
The numerical box of radiative transfer calculation is $r\le 100r_\mathrm{g}$ 
with $512\times 512$ grids.
Radiation transfer calculation is made for each radiation frequency, $\nu$.
\textcolor{black}{
The angular diameter of the M87 BH is $r_{\rm g} = 3.8 {\rm \mu as}$ for the distance of $D=16.7$ Mpc (\cite{Mei07}) and the mass of $M_{\rm BH} = 6.5 \times 10^9 M_\odot$ (\cite{EHT19a}).
}

\textcolor{black}{
Since the GRMHD simulations only give proton temperature ($T_{\rm p}$) distributions,
we need to prescribe how to determine electron temperatures, $T_{\rm e}$.
In the present study, we calculate the electron temperature 
by using the following relation (introduced by \cite{Mo16}),
\begin{equation}
	\frac{T_\mathrm{p}}{T_\mathrm{e}} = R_\mathrm{high}\frac{\beta^2}{1+\beta^2} + R_\mathrm{low}\frac{1}{1+\beta^2},
	\label{Te_rel}
\end{equation}
where $\beta$ is the ratio of gas pressure to magnetic pressure (so-called plasma-$\beta$)
and $R_{\rm high}$ and $R_{\rm low}$ ($\equiv 1$) are numerical constants.
Roughly, $R_\mathrm{high}$ corresponds to the ratio of proton to electron temperatures 
in the inner disk, whereas $R_\mathrm{low}$ corresponds to the same but in the jet region.
Each parameter value of calculated models in the present study will be summarized in section 2.3.
}

\begin{figure}
  \begin{center}
    \includegraphics[width=7cm]{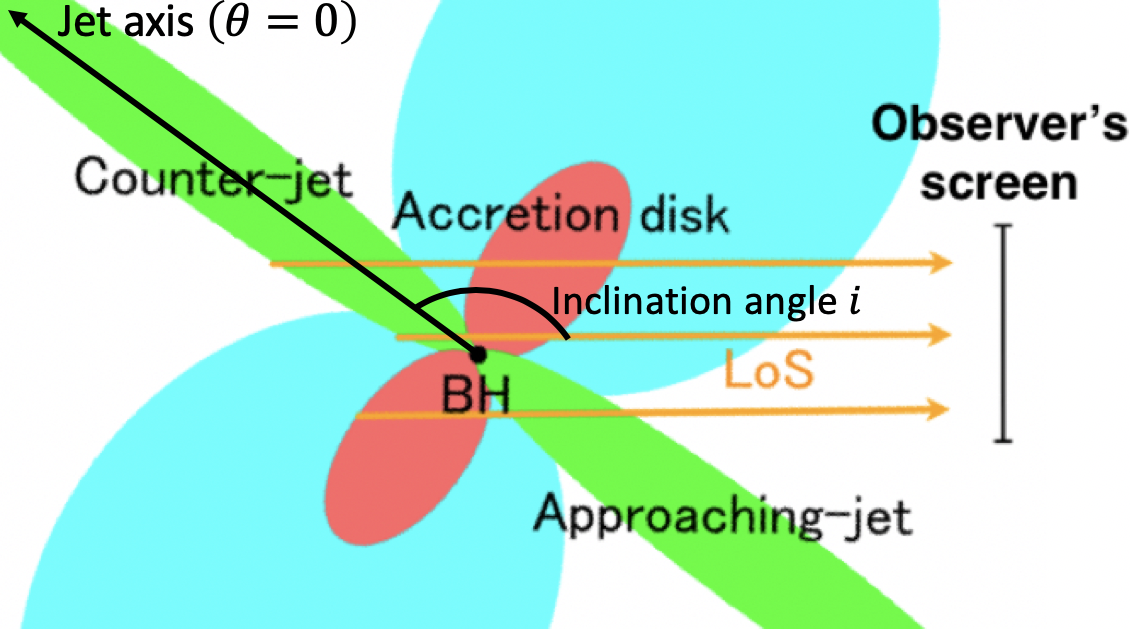}
  \end{center}
  \caption{A brief picture of our simulation. \textcolor{black}{Here we defined the inclination angle $i$ as angle between the `jet axis' (the line of $\theta=0$ in the coordinates) and the vector pointing from the origin to the center of screen.}
  	}
  \label{punch}
\end{figure}

\begin{figure*}
\begin{minipage}{0.48\hsize}
	\begin{center}
		\includegraphics[width=8cm]{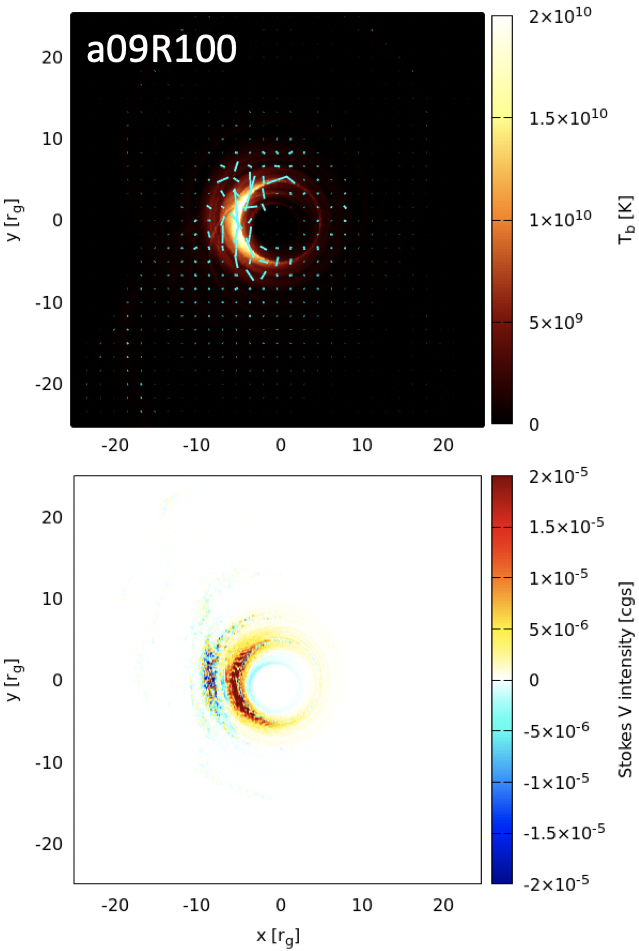}
	\end{center}
	\caption{(Top) The 230~GHz intensity map in brightness temperature $T_{\rm b} \equiv \left(2k_{\rm B}\nu^2/c^2\right)^{-1} I_\nu$ (color contours) for the fiducial model (Model a09R100) overlaid with the linear polarization vectors by EVPA (electric vector position angle) weighted with the linear-polarized intensity. (Bottom) The 230~GHz circular polarization (Stokes $V$, in cgs) image for the same model. 
	The intensity of $2 \times 10^{-5}$ in cgs at 230~GHz corresponds to the brightness temperature of $\approx 1.2 \times 10^9{\rm K}$.
	}
	\label{230_Rh100}
\end{minipage}
\hspace{5mm}
\begin{minipage}{0.48\hsize}
\vspace{-10mm}
\hspace{7.5mm}
	\begin{center}
		\includegraphics[width=7.9cm]{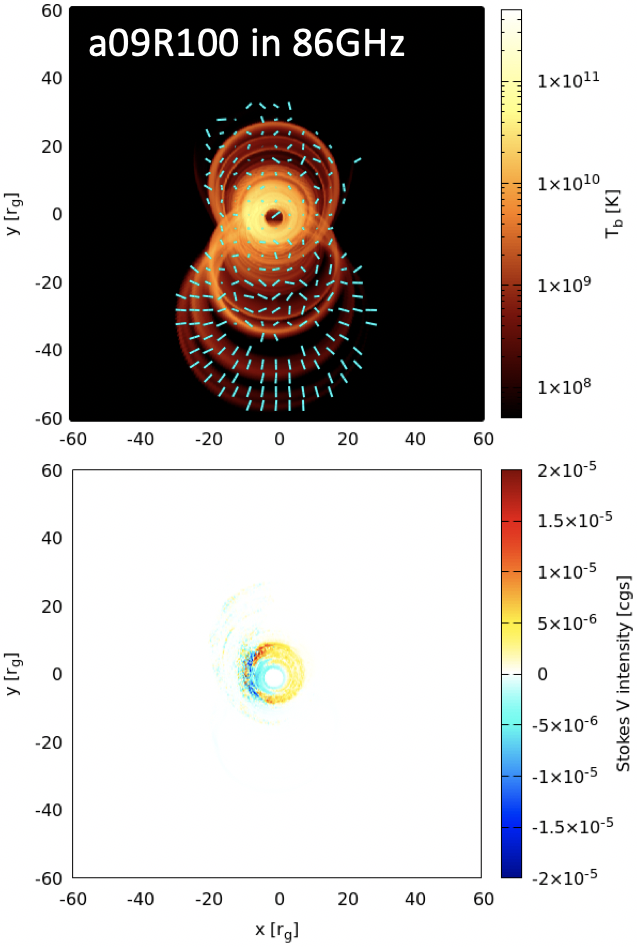}
	\end{center}
	\caption{Same as figure \ref{230_Rh100} but the images at 86~GHz.
 (Upper panel) The intensity map overlaid with the linear polarization vectors (not weighted),
of the central region of M87.
(Lower panel) The circular polarization (Stokes $V$) image.
Note that the box sizes of both panels are by a factor of 2.5 larger than those in figure \ref{230_Rh100}.
}
	\label{86_Rh100}
\end{minipage}
\end{figure*}

\subsection{Parameter setting for polarized GRRT calculation}\label{Cmodels}

There are many model parameters in the present study, but some of them are not free parameters (see also Table \ref{models} for calculated models).
We fix the value of $R_\mathrm{low} (=1)$
in addition to the distance, $D$, and the black hole mass, $M$, as are already mentioned,
while we vary the black hole spin, $a$,
inclination angle, $i$ (see Fig. 1), and the value of $R_\mathrm{high}$.

Bearing the M87 case in mind, we assign the fiducial parameter (or best-fit one based on our results) set as
$(a, R_\mathrm{high}, i) = (0.9 M_\mathrm{BH}, 100, \textcolor{black}{160\degree})$
(i.e., Model ``a09R100").
We should also note that the mass accretion rate ($\dot{M}$)
is not a free parameter but is determined so as to reproduce the observed flux of M87 core in 230~GHz to be $\simeq \textcolor{black}{0.5~\mathrm{Jy}}$ (\cite{EHT19d}) in the present study. 
\textcolor{black}{As a result}, the mass accretion rate \textcolor{black}{is distributed} in the range of $\textcolor{black}{(0.9 - 65) \times 10^{-3} M_\odot/\mathrm{yr}}$, depending not only on the black hole spin,
but also on the inclination angle (since the emission is highly anisotropic),
as is shown in table \ref{models}.

\section{Polarization Properties}
\subsection{Polarization images: fiducial model}

We first show in the top panel of figure \ref{230_Rh100} the $\nu =$ 230~GHz intensity (Stokes $I$, in the brightness temperature $T_{\rm b} \equiv \left(\frac{2k_{\rm B}\nu^2}{c^2}\right)^{-1} I_\nu$) image overlaid with the polarization vectors of our fiducial model. Here we observe the jet from below the equatorial plane ($i=160\degree$, see figure \ref{punch}), 
so the approaching jet (or counter jet) appears in the lower (upper) half of each panel.  
In the color contours we can observe the black hole ``shadow''. We also find
that the left half of a photon ring is brightened by special relativistic beaming effect due to plasma motion and gravitational blue- and red-shift due to black hole spin $a_\mathrm{BH}=0.9M_\mathrm{BH}$.

We calculated linear polarization vectors from the Stokes $Q$ and $U$ images at 230~GHz and overlay them on the color contours in the upper panel. Here the length of each vector is taken to be proportional to the polarized intensity ($\sqrt{Q^2+U^2}$). We see a rough tendency that
the polarization vectors are vertically ordered on its left side. In contrast, vectors in the outer region are disordered. 
We should note that linear polarization properties displayed in this figure do not directly reflect magnetic field structure, since they suffer Faraday rotation (discussed in \ref{Frot}).

We next show the circular polarization (Stokes $V$) image at 230~GHz in the bottom panel of figure \ref{230_Rh100}. The circular polarization component is large only in the side of the background counter jet, \textcolor{black}{and exhibits monochromatic, red (positive Stokes $V$) feature around the photon ring}. 

\citet{EHT19d} captured the 230~GHz horizon-scale image of M87 showing the black hole shadow without a clear detection of an extended jet structure due to a limited dynamic range of images achievable with the 2017 EHT array. Our fiducial model gives a consistent result with this observation.

\begin{figure*}
\begin{minipage}{0.5\hsize}
	\begin{center}
		\includegraphics[width=8cm]{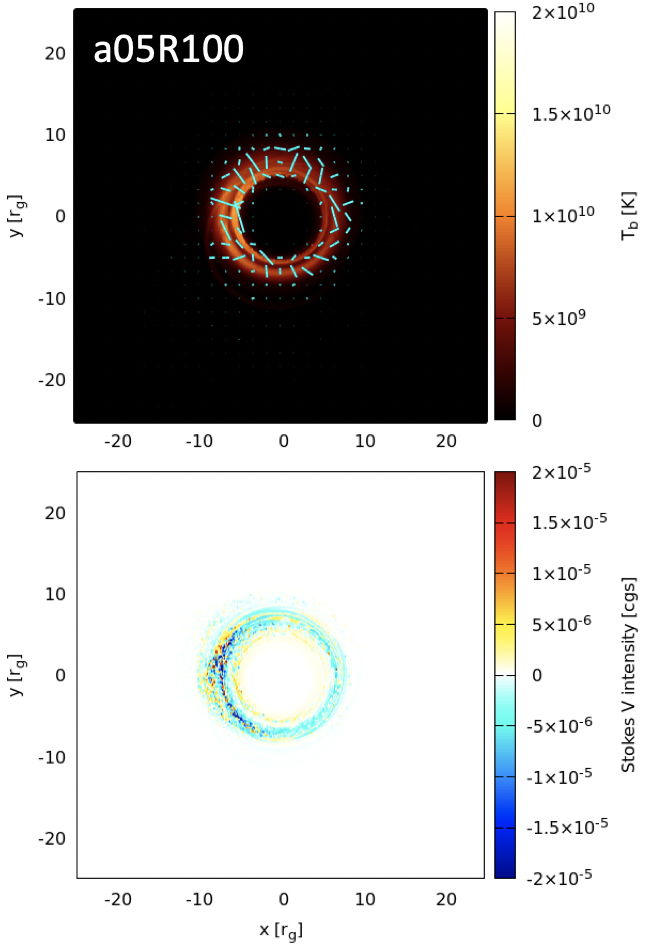}
	\end{center}
\end{minipage}
\begin{minipage}{0.5\hsize}
	\begin{center}
		\includegraphics[width=8cm]{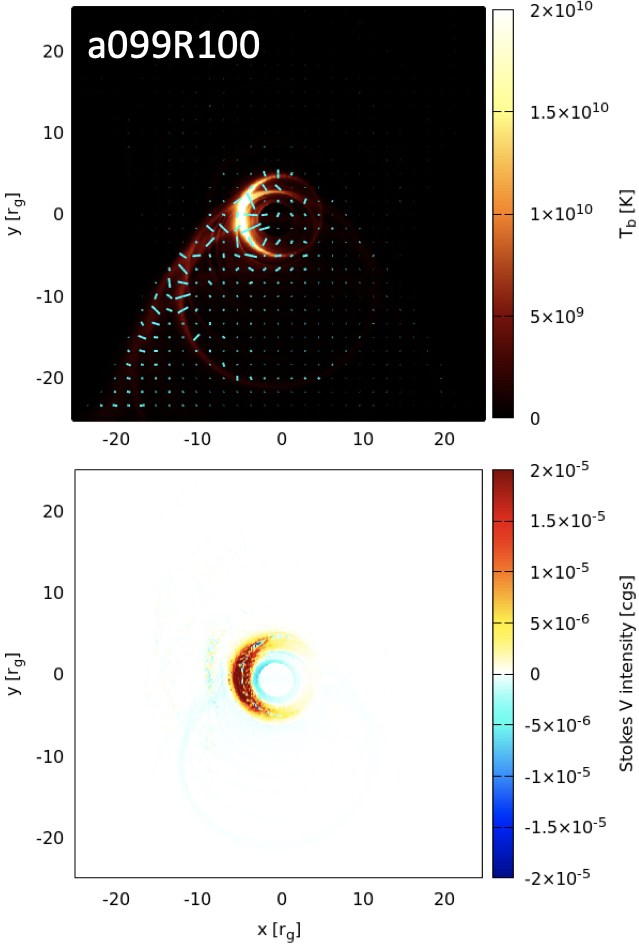}
	\end{center}
\end{minipage}
\caption{Same as figure \ref{230_Rh100} but for the low-spin model with $a = 0.5 M_{\rm BH}$ (left panels)
and for the high-spin model with $a = 0.99 M_{\rm BH}$ (right panels), respectively.}
	\label{230_a}
\end{figure*}

\begin{figure*}
\begin{minipage}{0.5\hsize}
	\begin{center}
		\includegraphics[width=8cm]{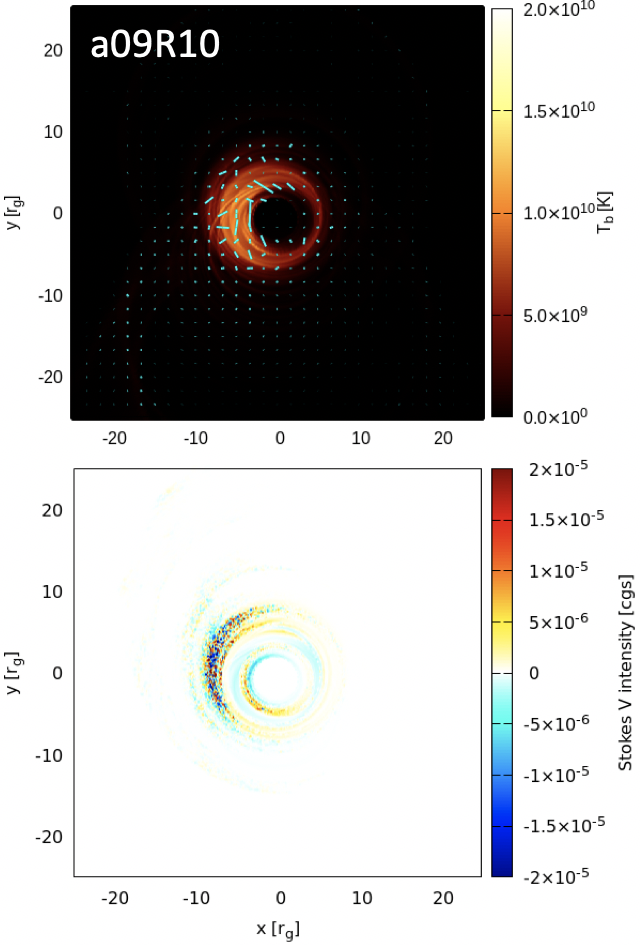}
	\end{center}
	\caption{Same as figure \ref{230_Rh100} but for the high-temperature disk ($R_{\rm high}=10$) model.}
	\label{230_Rh10}
\end{minipage}
\hspace{1mm}
\begin{minipage}{0.5\hsize}
	\begin{center}	
		\includegraphics[width=7.9cm]{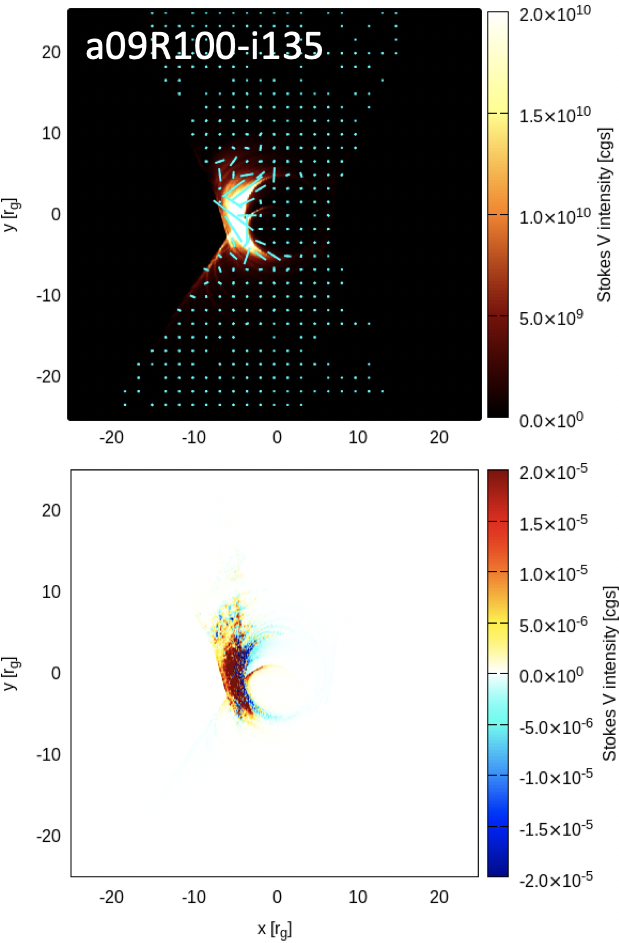}
	\end{center}
\vspace{-2mm}
	\caption{Same as figure \ref{230_Rh100} but for the nearly edge-on model with $i=135\degree$.
}
	\label{230_i135}
\end{minipage}
\end{figure*}

\subsection{Polarization images at 86~GHz}

We next show in figures \ref{86_Rh100} the intensity image overlaid with the vectors representing linear polarizations (upper panel) and the circular polarization image (lower panel) at 86~GHz.
Note that the spatial scale of this figure is by a factor of 2.5 as large as that of figure \ref{230_Rh100}.
We should also note in this figure that
the length of each vector in the upper panel is taken to be proportional to the linear polarization degree (i.e., $\sqrt{Q^2+U^2}/I$), not to the polarized intensity itself (i.e., $\sqrt{Q^2+U^2}$)
as is the case in figure \ref{230_Rh100}. 
This is to clearly display the polarization vectors in the jet regions with low intensity. 

In the upper panel of figure \ref{86_Rh100} we can clearly observe 
both of the approaching jet (in the lower part) and the counter jet (in the upper part).
This contrasts figure \ref{230_Rh100}, in which jet components are not clear at 230 GHz. 
We also notice that the both jets show limb-brightening.
This is because high Lorentz factors are achieved at the jet rim (see figure 12 of \cite{Na18}).
The counter jet is also bright, and its intensity looks comparable to that of the approaching jet.
This is partly because the gravitational lensing effects tend to enhance the light from the counter jet and partly because the bulk Lorentz factor is modest ($\Gamma_L \lesssim 3$) so that the boosting effects of the approaching jet cannot be so large \textcolor{black}{(see also  appendix \ref{86co} for discussion regarding the consistency of our calculation with the 86 GHz observation)}.

Let us next examine the linear and circular polarization properties in the upper and lower
panels of figure \ref{86_Rh100}, respectively.
We can see ordered linear polarization vectors in the outer, downstream region in the approaching jet, but they are less ordered in the region nearer to the black hole and in the counter jet.
\textcolor{black}{The well-ordered polarization-vector configuration in the outer region of our calculation image agree well with the results of 86 GHz polarimetry by Hada et al.~(2016). 
It is, however, premature to derive any useful constraints on theoretical models
from this comparison, since the knot is in the ambivalent location ($\sim 0.1\mathrm{mas} \approx 27r_{\rm g}$ from BH), where the polarization vectors exhibit a transition from well-ordered to disordered configurations. Furthermore, our simulation region ($\le 100 r_{\rm g}$) is smaller than the beam size of their observation and the size of the M87 core. In order to make thorough comparison with the polarimetry at 86 GHz and at even lower wavelengths we need larger simulation-box calculations, as well as observation with higher resolution and better sensitivity.}
We also see that the circular polarization is stronger in the counter jet than in the approaching jet.

\subsection{Polarization images: low- and high-spin models}

In this subsection we compare the results of the low spin model (a05R100) and of the high spin model (a099R100) with those of the fiducial model.
Figures \ref{230_a} are the same as figure \ref{230_Rh100}.but for the cases
with a lower spin of $a=0.5M_\mathrm{BH}$ (in the left panels)
and those with a high spin of $a=0.99M_\mathrm{BH}$ (in the right panels), respectively.
Note that the mass accretion rate is adjusted so as to give the same total intensity 
as that of the observation of M87 (see Table 1).

Let us first check the intensity profiles. 
We immediately notice much more symmetric ring shape in the low-spin model (see the upper left panel)
than in the fiducial model. This is partly because the beaming and de-beaming effects are not so large
and partly because jet acceleration is not so significant when the spin is small.
When the spin parameter is as large as $a=0.99M_\mathrm{BH}$, by contrast, the approaching jet image is clear 
(see the lower-left zone in the upper-right panel), which was not visible in the fiducial and low-spin models. 
This is because the plasma bulk motion is accelerated up to the Lorentz factor of $\Gamma_L\sim3$ when the spin is large,
so the approaching jet is more brightened than the counter jet by the beaming and de-beaming gap due to the bulk motion of plasmas in the jet.
Further, the toroidal motion of gas blobs gives the crescent-like image.
This demonstrate that the jet acceleration process strongly depends on the black hole spin (see figure 12 of \cite{Na18}).

We next examine the linear and circular polarization properties displayed in the upper and lower panels of figure \ref{230_a}.
We see that the low-spin model gives rise to a moderately ordered linear polarization vectors across the symmetric ring (at $r \sim 5-7 r_{\rm g}$ from the BH), while the high-spin model shows highly non-symmetric profile; relatively strong linear polarization is found in the left portions of the half-ring and in the region along the left edge of the approaching jet (see the lower-left region extending from the bright ring) due to the relativistic beaming effect. 
Circular polarization patterns in the low-spin model \textcolor{black}{exhibit monochromatic structure around the photon ring, with negative Stokes $V$},
which is common feature with the lower panel of figure \ref{230_Rh100}. 
Such circular polarization patterns \textcolor{black}{with positive Stokes $V$} are also seen in the high-spin model
(see the lower-right panel of figure \ref{230_a}).

\subsection{Polarization images: hot-disk model}
We next calculate the hot-disk model (Model a09R10) and show its intensity map overlaid with polarization vectors and circular polarization map in the upper and lower panels of figure \ref{230_Rh10}, respectively. 
We calculate the electron temperature according to equation (\ref{Te_rel}); that is, 
the smaller value of $R_\mathrm{high}$ means higher electron temperatures in the low-$\beta$ region;
i.e., the disk region, where thermal energy by far dominates magnetic energy. 
What happens, when the disk temperature is high?
Obviously, the radiation originating from the inner disk contributes much to the total flux,
leading to the enhancement of the wider image of the disk region in the intensity map (see the upper panel of figure \ref{230_Rh10}). 
As a result, the photon ring, which was clear in figure \ref{230_Rh100}, is no longer visible here. 
The linear polarization vectors are shorter (low polarization fraction) and show less-ordered structure. 
Interestingly, the circular polarization properties shown in the lower panel of figure \ref{230_Rh10} exhibit
\textcolor{black}{more noisy structure bicolored by red and blue (inconsistent in its sign) }.

\subsection{Polarization images: nearly edge-on model}

Finally, we calculate the 230~GHz polarization maps of the nearly edge-on model (Model a09R100-i135) with $i=135\degree$ and show the results in the upper and lower panels of figure \ref{230_i135}.

What are the most remarkable feature that arises by changing inclination angles? 
The straightforward answer to this question is that the beaming/de-beaming properties are distinct,
when we vary the inclination angle, since the relativistic Doppler boosting properties are very sensitive to the line-of-sight direction.
We show the linear and circular polarization properties of the low-inclination angle, nearly edge-on model
in the upper and lower panels in figure \ref{230_i135}, respectively, together with the intensity map.
The most notable feature in this figure is that the intensity contrast between 
the left and right sides of the photon ring is more enhanced when $i$ is large. 
Moreover, we can see the left-hand-side sheath of both of the approaching and counter jets. 
The linear polarization component is disordered in the brightest region, the left side of the photon ring. 
The circularly polarized light is strong in the left-hand side of this figure 
and in the counter jet region (in the upper part).

\section{Discussion}

\begin{figure*}
\begin{minipage}{0.5\hsize}
\hspace{15mm}
	\begin{center}
		\includegraphics[width=8cm]{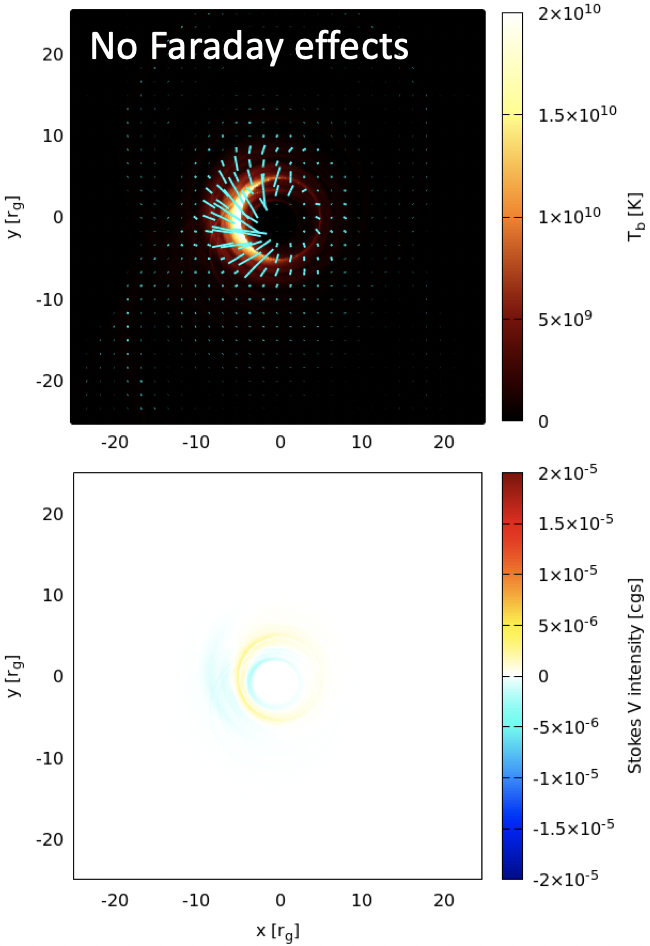}
	\end{center}
	\caption{Same as figure \ref{230_Rh100} but without Faraday effect. The observed frequency is 230~GHz.}
	\label{230_noF}
\end{minipage}
\hspace{5mm}
\begin{minipage}{0.5\hsize}
	\begin{center}
		\includegraphics[width=7.9cm]{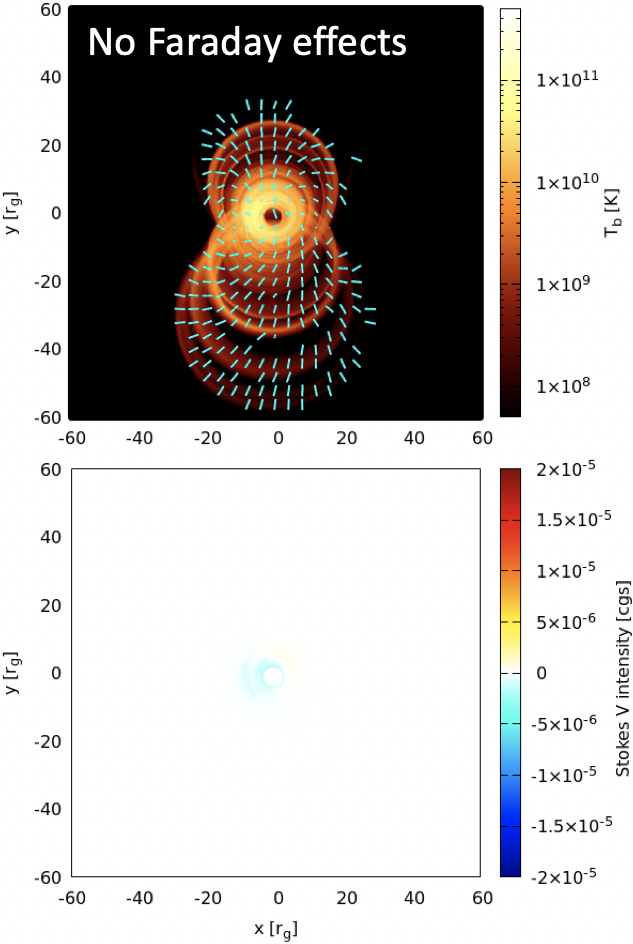}
	\end{center}
	\vspace{-1mm}
	\caption{Same as figure \ref{230_noF} but at the frequency of 86~GHz.}
	\label{86_noF}
\end{minipage}
\end{figure*}

\subsection{Faraday rotation and depolarization in images}\label{Frot}

The linear polarization properties of synchrotron emission can, in principle, convey the
information regarding the magnetic fields in the emitted region, however, 
the situation is not that simple, since the emission should have experienced the Faraday effects
when passing through magnetized plasma before reaching an observer. Therefore,
the polarization maps, such as those shown in figures \ref{230_Rh100} and \ref{86_Rh100},
may not simply reflect the magnetic structure in the AGN jet. 

In order to demonstrate that this is really the case, 
we performed {\lq\lq}fake{\rq\rq} ray-tracing calculations, in which we artificially eliminate
the Faraday effects by setting $\rho_Q=\rho_U=\rho_V=0$ in equation (\ref{poleq}) and show
the results in figures \ref{230_noF} and \ref{86_noF} for the observational frequencies at 230 GHz and at 86 GHz, respectively. 
These show the original polarization vectors; 
i.e., their directions are radial from the BH and thus across the photon ring, 
since the toroidal magnetic field components dominate near the black hole.
The light emitted from the region near the BH actually suffers from Faraday rotation 
when passing through magnetized plasma to go out and, hence, loses the original information 
regarding the magnetic fields, though we can still see a hint of the original
polarization vectors in the top panel of figure \ref{230_Rh100}.

In the 5th column of table \ref{models} we show the polarization fraction in the total flux 
at 230~GHz for each model. We notice in all models that
the polarization fraction is rather small, on the order of a few percents. 
That is, the total polarization fraction $\pi$ is suppressed, even though
local polarization fraction is as large as $\sim 10 \%$, since the polarization angle at each narrow area shows a diversity and strong polarizations are cancelled out in the total light in the case of nearly symmetric shape of bright regions. Conversely, asymmetrical images tend to give relatively high polarized fractions (see figure \ref{230_i135}).
Its top panel presents the lowest fraction \textcolor{black}{$0.82\%$} in our models, in spite of its concentrated feature. We interpret this to be a result of rather strong Faraday depolarization (see discussion in section \ref{polarimetry}).

At 86~GHz (figure \ref{86_Rh100} (top)), we also see that the linear polarization components from the inner flow and counter jet are largely depolarized (cf. figure \ref{86_noF}),
though there still remains certain amount of linear polarization in the outer approaching jet.
We should also note that these lights may suffer further depolarization in the region 
even outside our calculation box at $r>100r_g$.

\subsection{Comparison with the polarimetry}\label{polarimetry}

\citet{Kuo14} observed the core of M87 with the Submilli-meter Array at 230~GHz, aiming to constrain its mass accretion rate by the Faraday rotation measure (RM) measurements. 
They found its polarization fraction to be $\sim 1\%$ and 
evaluated the RM to be $-(2.1\pm1.8)\times 10^5~\mathrm{rad/m^2}$ with 
a 3$\sigma$ confidence range between $-7.5\times10^5~\mathrm{rad/m^2}$ 
and $3.4\times10^5~\mathrm{rad/m^2}$,
assuming that the Faraday rotation occurs in outer accretion flow rather than in jet interior, 
and also assuming a power-law density profile and equipartition between the magnetic fields and electrons, they limited the mass accretion $\dot{M}$ to be below 
$9.2\times10^{-4}M_\odot\mathrm{yr}^{-1}$ at $r=21r_S=42r_g$.

To compare our results with the polarimetry by \citet{Kuo14}, we calculated the total polarized flux by integrating the Stokes parameters on the screen at two wavelengths, 230 and 235~GHz, and calculated the polarization fraction $\pi$ and RM through the relationship of
\begin{equation}
 \mathrm{RM} = \frac{\chi_\mathrm{tot,1}-\chi_\mathrm{tot,2}}{{\lambda_1}^2-{\lambda_2}^2},
\end{equation}
where $\chi_\mathrm{tot}$ is the linear polarization degree of the total flux. We summarize
the calculation results for all models in the 5th and 6th columns of table \ref{models}.

There are some noteworthy features found in this table.
First, our fiducial model (a09R100) gives reasonable values of 
$\pi=2.3\%$ and $\mathrm{RM}=-2.9\times10^{5}\mathrm{rad/m^2}$,
which are fallen in the range given by the polarimetry.
In this sense, the fiducial model is consistent with the polarimetry observations. 

The mass accretion rate, $\dot{M}=1.4\times10^{-3}M_\odot/\mathrm{yr}$, is, however,
outside the allowed range estimated by \citet{Kuo14}. We interpret this discrepancy as a result of different magnetic field strengths and configurations adopted in their study and in ours. 
\citet{Kuo14} assumed that (1) the magnetic energy is equal to the internal energy of electrons (i.e., equipartition) and that (2) the field direction is radial, while we found in the GRMHD simulation data that (1) magnetic energy is significantly less than the electron energy and that (2) the field is mostly toroidal. We should caution that the Faraday rotation is proportional to $\int n \vec{B} d\vec{\ell}$ (where $n$ is the electron number density and integration is made along the line of sight vector, $\vec{\ell}$), and that we observe the M87 core from the high inclination angle, $i=160\degree$.
Therefore, they over-estimated the contribution of magnetic field to the RM, and so under-estimated the electron density and, hence, the mass accretion rate, compared with our evaluation.

Likewise, we find that models a09R10 and a099R100 are also consistent with the polarimetry, whereas the low spin model (a05R100) and nearly edge-on model (a09R100-i135) give much higher RM values by one and two orders of magnitude, respectively.
The high value of the low spin model could be associated with its high mass accretion rate,
$\textcolor{black}{6.5\times10^{-2}M_\odot/\mathrm{yr}}$, higher than the constraint by \citet{Kuo14} and 
also the values in other models, and this high $\dot M$ value could be due to
weaker acceleration by weak magnetically driven outflow and jet. 
High accretion rates means high densities of accretion flow in the region near to the BH, 
leading to enhancement of the Faraday rotation effects. 
The high value of nearly edge-on model (a09R100-i135),
$\textcolor{black}{1.1\times10^7~\mathrm{rad/m^2}}$ can be understood in terms of the orientation effects.
That is, the rays which reach us should have passed through the dense and turbulently magnetized flow near the equatorial plane and thus have experienced the Faraday effects more strongly
so that they should be extremely depolarized, compared with other cases with $i = 160\degree$. 

\textcolor{black}{
\subsection{Amplification mechanism of circular polarization}
}

\begin{figure*}[]
\begin{minipage}{0.48\hsize}
	\begin{center}
		\includegraphics[width=10cm]{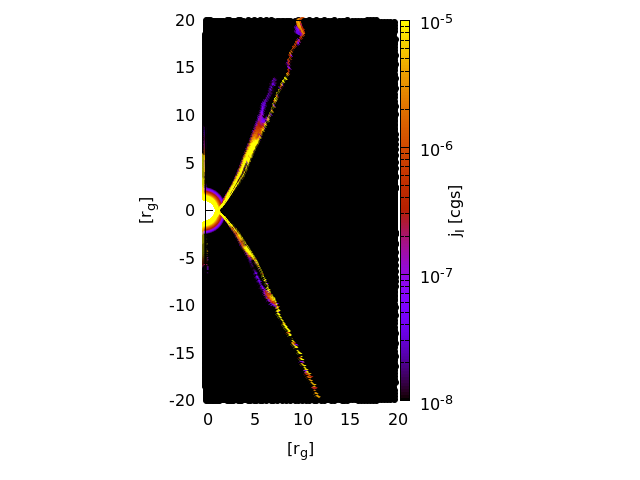}
	\end{center}
\end{minipage}
\hspace{-20mm}
\begin{minipage}{0.48\hsize}
	\begin{center}
		\includegraphics[width=10cm]{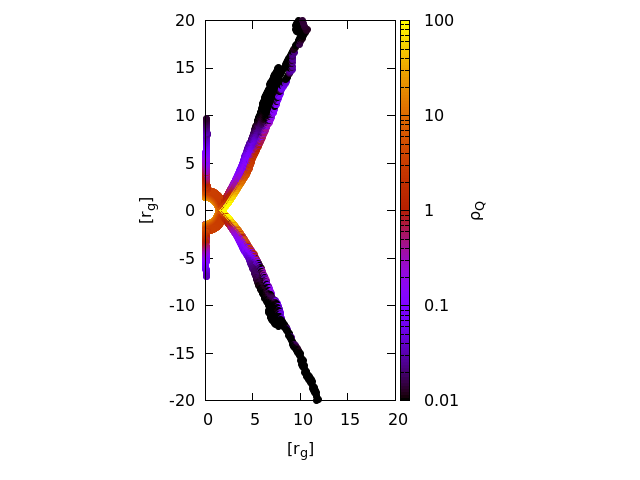}
	\end{center}
\end{minipage}
	\caption{Left: map of synchrotron emissivity (per $r_{\rm g}$) in 230~GHz estimated from magnetic strength and electron density and temperature without the effect of angle between light and field and relativistic effect by bulk motion of plasma, focused on the region near the black hole ($\sim 20r_{\rm g}$). Right: map of the Faraday conversion coefficient (per $r_{\rm g}$) in 230~GHz estimated as emissivity in the right. Only the region where $\rho_Q/\rho_V>10^{-2}$ is plotted.
	}
	\label{R100}
\end{figure*}

\textcolor{black}{
We see in all models bright area in the circular polarization image of the counter-jet. Especially in models with low-temperature disk (Models a09R100, a05R100, a099R100), we find a uniform, ring-like structure around the photon ring (see lower panels of figures \ref{230_Rh100} and \ref{230_a}). 
In these images \textcolor{black}{the fraction of circular polarization is $\gtrsim 10\%$, much higher than the original fraction ($\lesssim 1\%$) expected in the synchrotron process}. We thus conclude that such high circular polarization degree should be a result of the Faraday conversion from the linear polarization produced by the synchrotron emission. This conclusion is supported by figures \ref{230_noF} and \ref{86_noF}, in which \textcolor{black}{the Faraday effects are removed by hand (setting $\rho_Q = \rho_U = \rho_V = 0$)}, and which show rather low circular polarization.
}

\textcolor{black}{
We need to make caution that such large circular polarization can be produced,
if and only if both of synchrotron emission and Faraday effects occur nearly
within the same place which is optically thin \textcolor{black}{for synchrotron self-absorption}, Faraday thick, and threaded with ordered magnetic fields.
We will explain the reason for this in the following.
}

\textcolor{black}{
Let us first note that strong Faraday conversion is necessary, but is not sufficient, for producing large circular polarization. This is because the Faraday conversion not only causes the conversion from linear polarization to circular polarization, but also yields its back reaction; i.e., conversion from circular polarization to linear polarization. In the limit of very strong Faraday conversion, therefore, both of linear and circular polarization degrees should oscillate from positive to negative and from negative to positive, as radiation propagates (cf. appendix A of \cite{Mo18}). 
Such features are not seen, at least in the circular polarization maps of the low-temperature disk models.
Then, how is such uniform amplification of circular polarization realized?
It will be instructive in this respect to examine the discussion by \citet{De16} who considered an equation with uniform emission, Faraday rotation, and conversion (without absorption) in his appendix C, 
\begin{equation}\label{j_rho}
	\frac{d}{ds}\left(
		\begin{array}{c}
			Q \\
			U \\
			V \\
		\end{array}
		\right)
		=
		\left(
		\begin{array}{c}
			j_Q \\
			j_U \\
			j_V \\
		\end{array}
		\right)
		-
		\left(
		\begin{array}{cccc}
			 0 & \rho_V & 0 \\
			 -\rho_V & 0 & \rho_Q \\
			 0 & -\rho_Q & 0 \\
		\end{array}
		\right)
		\left(
		\begin{array}{c}
			Q \\
			U \\
			V \\
		\end{array}
		\right),
\end{equation}
where \textcolor{black}{$j_Q, j_U, j_V, \rho_Q, \rho_V$ are the same as those in eqs. (8)-(11),  $s$ is the distance, and $\theta_B$ is an angle between the magnetic field line and the propagation vector of light-ray (line of sight)}. Here we can assume $\rho_U=0$ without loss of generality. These equations have analytic solutions,
\begin{equation}\label{an}
	Q(s) = \frac{\rho_Q}{\rho^2}(j_Q\rho_Q+j_V\rho_V)s - \frac{\rho_V}{\rho^3}(j_V\rho_Q-j_Q\rho_V)\mathrm{sin}\rho s - \frac{j_U\rho_V}{\rho^2}(1-\mathrm{cos}\rho s ),
\end{equation}
\begin{equation}
	U(s) = \frac{j_U}{\rho}\mathrm{sin}\rho s + \frac{j_Q\rho_V-j_V\rho_Q}{\rho^2}(1-\mathrm{cos}\rho s ),
\end{equation}
\begin{equation}
	V(s) = \frac{\rho_V}{\rho^2}(j_Q\rho_Q+j_V\rho_V)s - \frac{\rho_Q}{\rho^3}(j_Q\rho_V-j_V\rho_Q)\mathrm{sin}\rho s + \frac{j_U\rho_Q}{\rho^2}(1-\mathrm{cos}\rho s ),
\end{equation}
here $\rho\equiv \sqrt{\rho_Q^2+\rho_V^2}$.
}

\textcolor{black}{
We, here, consider the amplification of $V(s)$ under the condition of $|j_Q|, |j_U| \gg |j_V| \sim 0$. 
If the path is Faraday thin, $\rho s \ll 1$, the solutions written up to second order of $\rho s$ are
\begin{displaymath}
 Q(s)\simeq j_Qs - \frac{j_U}{2\sqrt{2}\rho}(\rho s)^2,
\end{displaymath}
\begin{equation}
 U(s) \simeq j_Us +\frac{j_Q}{2\sqrt{2}\rho}(\rho s)^2,
\end{equation} 
\begin{displaymath}
 V(s) \simeq \frac{j_U}{2\sqrt{2}\rho}(\rho s)^2,
\end{displaymath}
(Here, we set $\rho_Q = \rho_V = \rho/\sqrt{2}$ for simplicity.)
We thus understand that $V(s)$ actually increases but not faster than $Q(s)$ and $U(s)$, as long as the path length ($s$) is short. 
}

\textcolor{black}{
If the path is Faraday thick, $\rho s \gg 1$, conversely, 
the first term of $Q(s)$ and $V(s)$, which is linear with respect to the path length $s$, dominates over other terms, and also over the terms in $U(s)$. 
We thus have
\begin{equation}
  Q(s) \sim \frac{1}{2}\frac{\rho_Q^2}{\rho^2} j_Qs,\quad
  U(s) \sim 0,\quad
  V(s) \sim \frac{1}{2}\frac{\rho_Q\rho_V}{\rho^2} j_Qs
\end{equation}
\textcolor{black}{from eq. (\ref{an}) - (20),} \textcolor{black}{by eliminating the oscillating terms, which should be negligible in the limit of large Faraday optical depth, $\rho s \gg 1$.} (Here, we still assume $|\rho_Q| \sim |\rho_V|$ but distinguish $\rho_Q$ and $\rho_V$ for the convenience of later discussion.)
As a result, circular polarization degree can grow up to become comparable to that of linear polarization in the condition of balanced Faraday effects,
provided that $|\rho_Q| \sim |\rho_V|$ and that Faraday optical depth is large, $\rho s \gtrsim 1$. This occurs even with negligible circular-polarized emission ($|j_V| \ll |j_Q|, |j_U|$).
}

\textcolor{black}{In detail, the sense of amplification (positive or negative) is associated with that of Faraday rotation, which is proportional to the line-of-sight component of magnetic field ($\rho_V \propto {\rm cos}\theta_B$), and suggests the `direction' (approaching to or away from us) of magnetic field line.}

\textcolor{black}{
The biggest assumptions made in the above analysis resides in uniform Faraday coefficients; in other words, magnetic field lines are assumed to be well-ordered and their strengths, as well as electron temperatures, do not vary so much from the place of synchrotron emission to that of Faraday conversion. Otherwise, $\rho_Q$ or $\rho_U$ may change their sign, leading to reduction of the circular polarization degree. Therefore, 
there are three conditions for the growth of circular polarization: (1) the synchrotron plasma is optically thin but Faraday thick. (2) The Faraday rotation and conversion are balanced there; i.e., $|\rho_Q| \sim |\rho_V|$. (3) Synchrotron emission and Faraday conversion take place roughly in the same region threaded with well-ordered magnetic fields and with nearly uniform electron temperatures.
}

\textcolor{black}{
Let us apply this analysis to our models and search for the regions where the three conditions mentioned above are met.
In short, the above three conditions are satisfied in the jet 
rim near the black hole but only in models with low-temperature disk.
We first show the place where synchrotron radiation originates from in the left panel of figure \ref{R100}. Here we plot the contours of synchrotron emissivity at 230~GHz for our fiducial model. (Note that synchrotron emissivity in this plot is estimated simply from magnetic-field strength, electron density in the fluid rest flame, and electron temperature at that point and that relativistic effects are not considered.)
\textcolor{black}{Although no sigma-cutoff condition is adopted here,} we find that the emission region is concentrated at the jet rim below $z \sim 8 r_{\rm g}$, as is indicated by the yellow color.
\textcolor{black}{The reason of why no emission is found within the funnel region is that the most luminous region in our model nearly coincides with the funnel wall, where $\sigma \sim 1$, and not the region with high $\sigma$ values.
Note that the same feature is found in other SANE-jet model (see, e.g., Figure 12 of \cite{Mo16}).}
It is important to note that contribution from the disk (inflow) is negligible
in this model, since the disk temperature is relatively low ($R_\mathrm{high}=100$).
}

\textcolor{black}{
Next, let us examine the place where the Faraday effects are large. 
We plot such regions in the right panel of figure \ref{R100},
finding that the region similar to the yellow region in the left panel
shows large value in the right panel; more precisely,
the jet rim but below $z \sim 5 r_{\rm g}$ shows high conversion.
Thus, the high emissivity region and high conversion region are nearly identical.
Furthermore, we confirm that the condition of $|\rho_Q| \sim |\rho_V|$ also holds in this rim  region.
This is reasonable, given that the ratio of coefficients of the Faraday effects is $|\rho_Q/\rho_V| \propto BT_e^3$ in the limit of high temperature \textcolor{black}{(because we obtain $\rho_Q \propto B^2(\theta_e+K_1(\theta_e^{-1})/6K_2(\theta_e^{-1})) \sim B^2T_e$ and $\rho_V \propto BK_0(\theta_e^{-1})/K_2(\theta_e^{-1}) \sim BT_e^{-2}$ (\cite{Sh08}; \cite{De16}) in the limit of high-temperature, $\theta_e \equiv k_{\rm B}T_e/m_ec^2 \gtrsim 1 \Leftrightarrow T_e \gtrsim 10^{10}{\rm K}$, where $K_\alpha(x)$ is $\alpha$-rank modified Bessel function)}, so the hotter and magnetically stronger, the higher this ratio.
Furthermore, we find that only the region of near the black hole ($\lesssim 5r_g$) is Faraday thick, where toroidal magnetic fields are dominant and well-ordered toroidal fields drive the jet. 
}

\begin{figure*}[]
\begin{minipage}{0.48\hsize}
	\begin{center}
		\includegraphics[width=10cm]{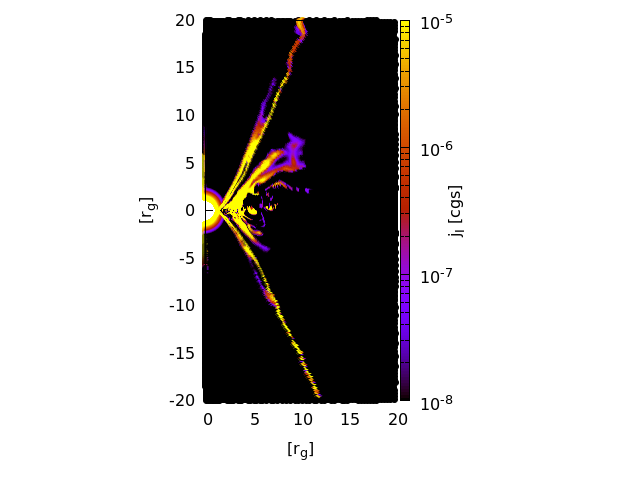}
	\end{center}
\end{minipage}
\hspace{-20mm}
\begin{minipage}{0.48\hsize}
	\begin{center}
		\includegraphics[width=10cm]{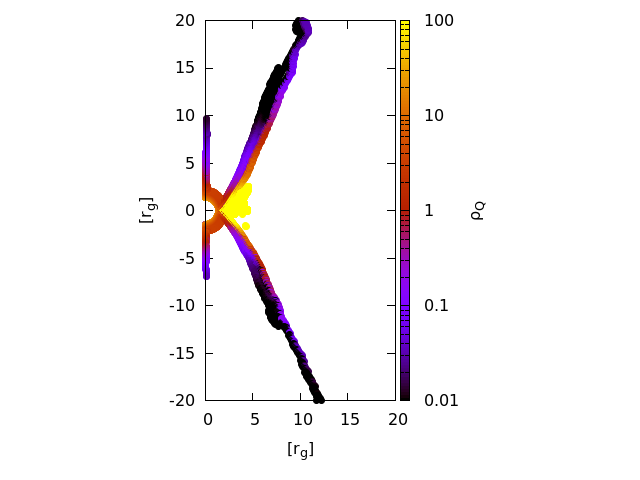}
	\end{center}
\end{minipage}
	\caption{Same maps as figure \ref{R100}, but for the hot disk model (a09R10).
	}
	\label{R10}
\end{figure*}

\textcolor{black}{
To conclude, all the conditions for the amplification of circular polarization is satisfied in the jet rim near the black hole, 
Hence, radiation originating from that region undergoes significant amplification 
of circular polarization before reaching an observer, 
thus producing ring-like circular polarization image.
In other huge regions with large Faraday rotation
circular polarization does not grow appreciably (cf. equation (A5) of \cite{Mo18}). 
\textcolor{black}{Focusing on the sign of amplified circular polarization, we can state that the low spin model a05R100 has magnetic field with different direction from other two models a09R100 and a099R100, as described above.}
}

\textcolor{black}{
We finally consider the case of high-temperature disk (a09R10), which exhibits rather noisy and disordered structure in circular polarization image (see Fig. \ref{230_Rh10}). We can see that both of the emission region and Faraday region are larger, compared with those of low-temperature models shown in figures \ref{R100}, and extend to the inner disk. This is because the inflow part, which is in the form of RIAF (radiatively inefficient accretion flow; see, e.g., \cite{KFM08}, chap. 9), is now visible. Since the RIAF is threaded with turbulent magnetic field lines, \textcolor{black}{the conditions of ordered field} described above are not satisfied there. \textcolor{black}{To be more precise, high temperature, which gives rise to large absolute values of the conversion coefficients ($\rho_Q$ and $\rho_U$), is not sufficient to enhance CP components by conversion. In addition, magnetic field lines should have ordered structure in the whole conversion region, since otherwise the CP component disorderly varies, changing their sign and absolute values, along each path, so that the integral along the path will get smaller. 
This is the reason why the CP image of hot disk model (with high turbulence and high temperature) shows chaotic structure with diverse degrees of zero up to $\sim \pm10\%$, while the CP image of fiducial model (with low turbulent and low temperature) exhibits uniform structure in its sign with the degrees of $\gtrsim 10\%$.}
We can thus understand why high-temperature disk model yields not so large circular polarization.
}

\textcolor{black}{
In summary, we specify the conditions that circular polarization grows up through the Faraday effects by ordered magnetic field. As a consequence, it will be possible to elucidate the magnetic field \textcolor{black}{direction} and temperature distribution of emission plasma, from the observation of circular polarization. It may also be possible to identify the emission regions, either of disk or jet, or both. \textcolor{black}{Comparing the SANE and MAD regimes in terms of circular polarization, we can infer that Faraday conversions would be stronger in the semi-MAD case than in the SANE case. This is because stronger and better-ordered magnetic field lines make the conversion process more effective.}
}

\subsection{Comparison with previous studies}

\citet{Mo17} calculated a polarized radiative transfer for the SANE (standard and normal evolution) models of M87 for the spin parameter of $a_\mathrm{BH}\approx0.94 M_\mathrm{BH}$, and the inclination angle of $i=20\degree$.
[Note that the base GRMHD models of our calculation are classified to the semi-MAD (magnetically arrested disk) model.]
A big distinction between their results and ours appears in the dependence of the estimated mass accretion rate on the changes of electron temperatures in the disk (or the $R_{\rm high}$ value, see equation \ref{Te_rel}). 
According to their table 1 $\dot{M}$ varies from $9\times10^{-3} M_\odot/\mathrm{yr}$ to $1\times10^{-3} M_\odot/\mathrm{yr}$ as $R_\mathrm{high}$ changes from 100 to $10$, while  $\dot{M}$ only changes by a factor of $\sim 1.5$ (from \textcolor{black}{$1.4\times10^{-3} M_\odot/\mathrm{yr}$ to $9.0\times10^{-4} M_\odot/\mathrm{yr}$}) for the same change in $R_{\rm high}$ in our case.
This is because the semi-MAD jet is more powerful than the SANE jet that the contribution of 
the jet emission relative to the disk emission should be much larger. As a result, the disk temperature is less important in our calculations, than in theirs, when calculating the total flux.

This different dependence also affects the observational properties of the linear polarization vectors. 
Their figures 6 and 7 show that the original polarization structure survives to some extent 
in a hot-disk models (their RH1 and RH5), while it is disordered and depolarized in a cold-disk models (RH20 and RH40). In our simulations, by contrast, Faraday rotation and depolarization are fatal even in the hot-disk model (a09R10) due probably to
the strong magnetic field in the semi-MAD scheme and due also to the axial symmetry of the GRMHD simulation which we adopted.

As for the Faraday RMs, our absolute values in Models a09R100, a09R10, and a099R100 are in the same order as theirs. 
In addition, it might be noted that their circular polarization fraction is comparable to the linear one in all models, which is the common feature also in our calculations.

\section{Conclusion}

We performed general relativistic, polarimetric radiative transfer calculations
based on the axisymmetric GRMHD simulation data by \citet{Na18}
and made polarization images in the event-horizon scale to compare with future EHT polarimetry to elucidate the roles of magnetic fields for launching jets and outflow.
Our results can be summarized in the following way:
\begin{itemize}
\item 
The calculated images at 230~GHz are sensitive to the black hole spin. 
Our fiducial model with the black hole spin of $a_\mathrm{BH}=0.9M_\mathrm{BH}$, 
can reproduce the asymmetric crescent-shape photon ring and absence of jet features
as are observed in the recent EHT observations of M87.

\item 
We calculated linear and circular polarization maps.
We found in all models that the linear polarization vectors, 
that can be seen at a distant observer, 
could be significantly modified by the Faraday rotation and depolarized
so that the original information regarding the magnetic field properties may be lost
at least partly. \textcolor{black}{We also find that the circular polarization can grow via Faraday conversion of the linear polarization in the hot plasmas threaded by ordered magnetic fields, thus the circular polarization images provide information regarding the direction of magnetic field lines.
}

\item
We compare our results with the polarimetry of M87 core at 230~GHz
in terms of rotation measure (RM) and the estimated mass accretion rate, 
finding that the fiducial model with $a = 0.9 M_{\rm BH}$
are again favored over other models with $a = 0.5 M_{\rm BH}$ or $0.99 M_{\rm BH}$.

\item One of the most outstanding issues is to compare the polarization images 
obtained by highly-resolved polarimetry of M87 and LLAGNs, 
to elucidate the magnetic field structures and the origins of jet eruption near the SMBH.

\end{itemize}

\bigskip

\section*{Acknowledgement}

We wish to acknowledge Masanori Nakamura for provision of GRMHD simulation data sets and stimulating discussions. 
This work is supported in part by JSPS Grant-in-Aid for Scientific Research (A) (17H01102 KO), same but for Scientific Research (C) (17K0583 SM, 18K03710 KO), and for Scientific Research on Innovative Areas (18H04592 KO), and JSPS KAKENHI Grant Number JP18K13594 (TK).
This research is also supported by MEXT as a priority issue (Elucidation of the fundamental laws and evolution of the universe) to be tackled by using post-K Computer and JICFuS.
KA is a Jansky Fellow of the National Radio Astronomy Observatory, and financially supported in part by a grant from the National Science Foundation (AST-1614868).
The National Radio Astronomy Observatory is a facility of the National Science Foundation operated under cooperative agreement by Associated Universities, Inc. 
Numerical computations were in part carried out on Cray XC30 and XC50 at Center for Computational Astrophysics, National Astronomical Observatory of Japan. 
Numerical analyses were in part carried out on analysis servers at Center for Computational Astrophysics, National Astronomical Observatory of Japan.

\newpage
\onecolumn

\appendix

\setcounter{section}{0} 
\renewcommand{\thesection}{\Alph{section}} 
\setcounter{equation}{0} 
\renewcommand{\theequation}{\Alph{section}\arabic{equation}}
\setcounter{figure}{0} 
\renewcommand{\thefigure}{\Alph{section}\arabic{figure}}
\setcounter{table}{0} 
\renewcommand{\thetable}{\Alph{section}\arabic{table}}

\textcolor{black}{
\section{Performance tests}\label{tests}
}

\begin{figure*}
	\begin{center}
		\includegraphics[width=16cm]{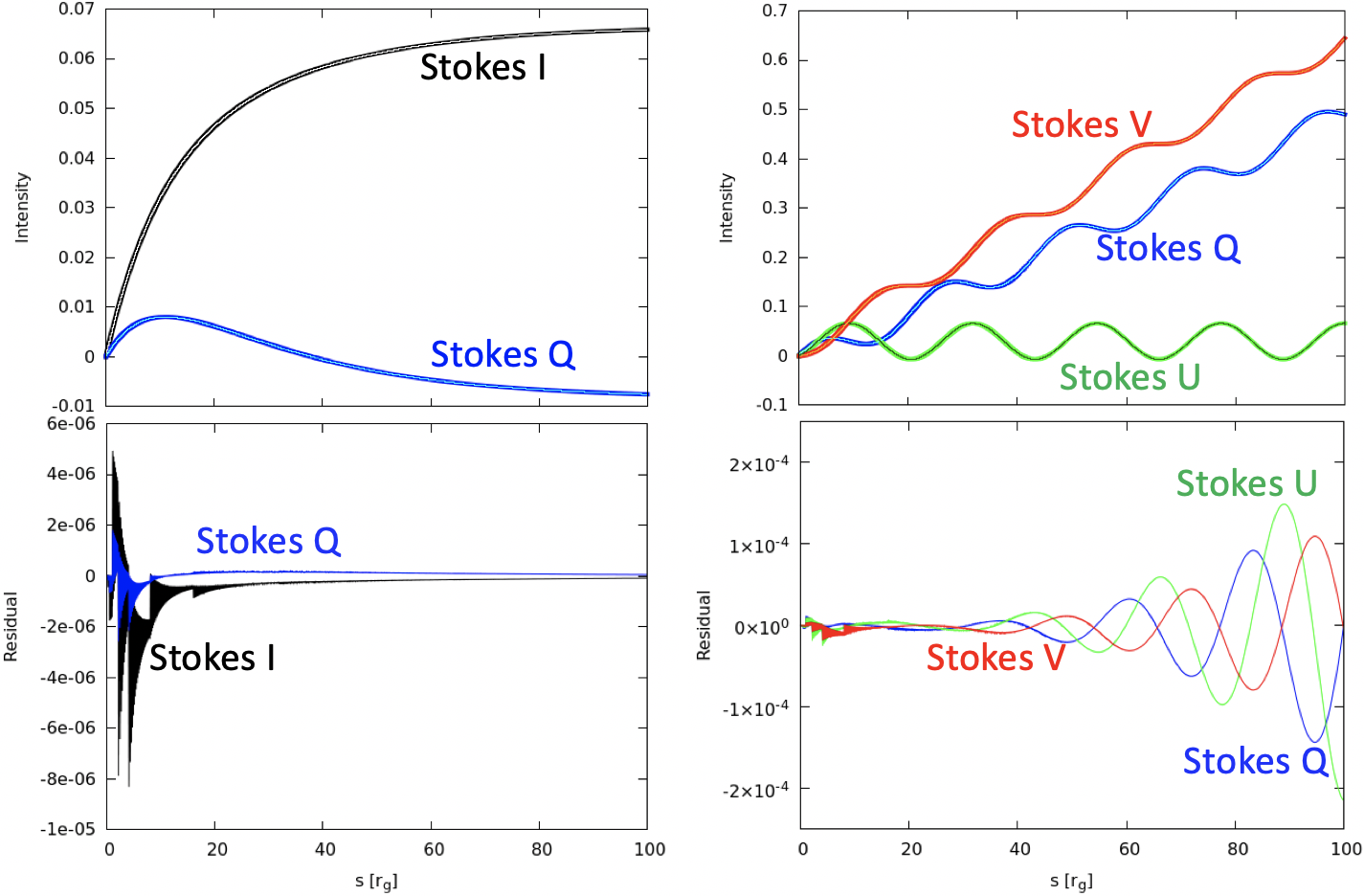}
	\end{center}
	\caption{Top left: Evolution of Stokes $I$ and $Q$ for a test problem of equations (\ref{test1}), with numerical solutions (bold) by our code and analytic ones (fine). Bottom left: Evolution of each of residual for equations (\ref{test1}). Top right: Evolution of Stokes $(Q,U,V)$ for a test problem of equations (\ref{j_rho}), as the top left. Bottom right: Evolution of residuals for equations (\ref{j_rho}).
	}
	\label{test}
\end{figure*}

\begin{figure*}[]
	\begin{center}
		\includegraphics[width=12cm]{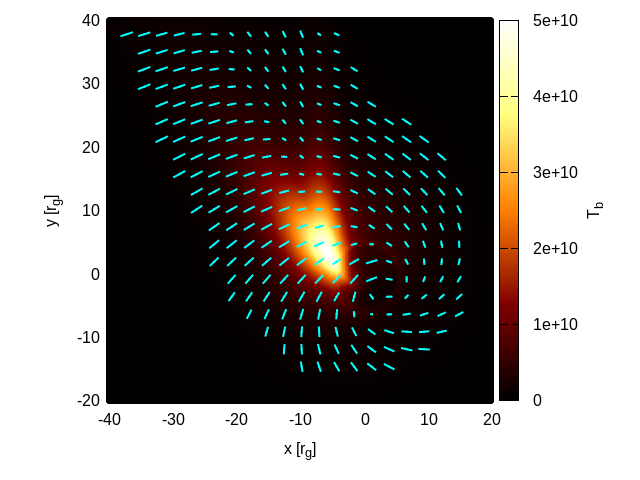}
	\end{center}
	\caption{345~GHz linear polarization map (brightness temperature in color contour, in linear scale, and polarization vectors by EVPA in ticks) of semi-analytical force-free jet model for M87 by \citet{BL09}. This corresponds to their figure 7 (for their model M0) and figure 6 of \citet{De16}.
	}
	\label{BL09}
\end{figure*}

\textcolor{black}{
Here we performed polarized radiative transfer simulations with a new code, which we developed, following \citet{De16}.
To evaluate the performance of the code, we did the same tests as those employed by \citet{De16}. First, we numerically solved the transfer equations, assuming uniform emission and absorption for Stokes $I$ and $Q$ components (see his appendix C).
\begin{equation}\label{test1}
	\frac{d}{ds}\left(
		\begin{array}{c}
			I \\
			Q \\
		\end{array}
		\right)
		=
		\left(
		\begin{array}{c}
			j_I \\
			j_Q \\
		\end{array}
		\right)
		-
		\left(
		\begin{array}{cccc}
			 \alpha_I & \alpha_Q \\
			 \alpha_Q & \alpha_I \\
		\end{array}
		\right)
		\left(
		\begin{array}{c}
			I \\
			Q \\
		\end{array}
		\right),
\end{equation}
and compared the results with the analytic solutions, [(C2) and (C3) in \citet{De16}], in the left panels of figure \ref{test}. We find good agreement between them.
}

\textcolor{black}{
Second, we compared the numerical and analytic solutions of equation with uniform polarized emission and Faraday effects for Stokes $(Q, U, V)$, which is the same as equation (\ref{j_rho}) in the present paper [cf. (C5-7) in \citet{De16}], 
and showed them and its residues from the analytical results (see equation (\ref{an})) in the right panels of figure \ref{test}.
Although the residues by our code are larger than those of \cite{De16} in two orders of magnitude, these are small enough for our discussion here.
}

\textcolor{black}{
In addition, we made a 345 GHz linear polarization map of semi-analytical force-free jet model introduced by \citet{BL09}, and show it in figure \ref{BL09}[, which corresponds to their figure 7 (M0) and figure 6 of \citet{De16}]. We find good agreement between them, in terms of asymmetric, branches-like brightened structure by helical bulk motion of plasma, and well-ordered polarization vectors.}

\newpage

\setcounter{equation}{0}
\setcounter{figure}{0}

\begin{figure*}[]
	\begin{center}
		\includegraphics[width=12cm]{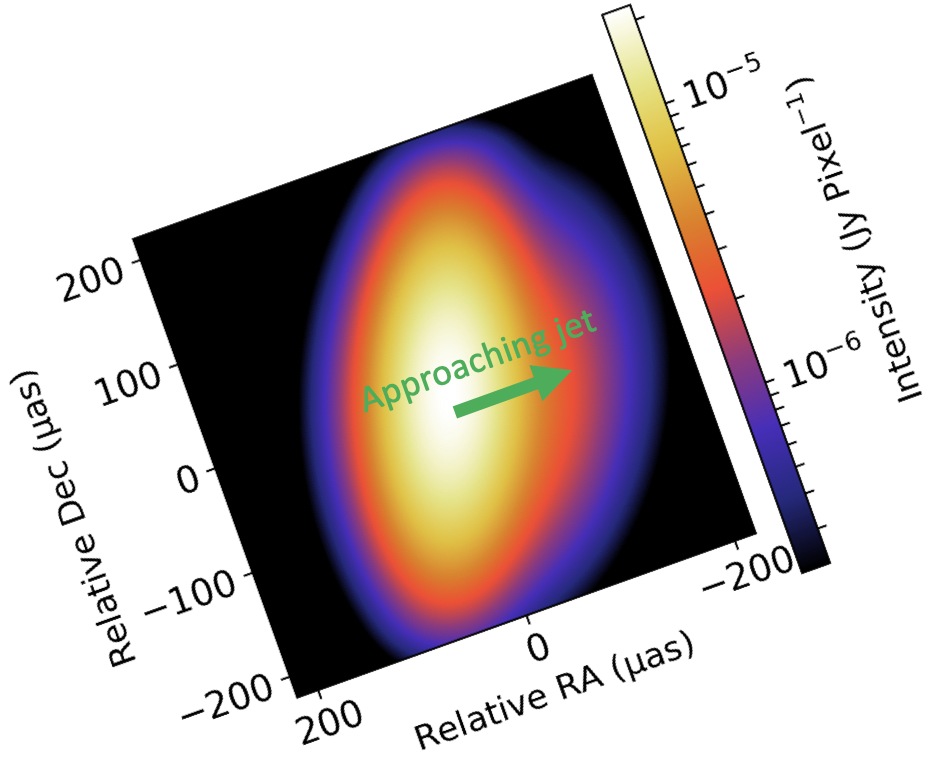}
	\end{center}
	\caption{\textcolor{black}{86~GHz intensity image of our fiducial model (cf. figure \ref{86_Rh100}) in log scale, convolved with Gaussian beam of $\approx 70{\rm \mu as}\times 200{\rm \mu as}$ inclined $-20\degree$ to the jet axis.
	}}
	\label{86con}
\end{figure*}

\textcolor{black}{
\section{Convolved image in 86~GHz}\label{86co}
}
\textcolor{black}{
The 86~GHz image in figure \ref{86_Rh100} might seem to contradict the well-known feature of M87 jet, an outstanding approaching-component as seen in larger-scale observations (cf. \cite{Ha16}; \cite{Ki16}; \cite{Wa18}). To dispel this doubt, we post-processed 86~GHz image of the fiducial model (the brightness contour of upper panel in figure \ref{86_Rh100}) with a python-interfaced library `SMILI' (Sparse Modeling Imaging Library for Interferometry, \cite{Ak17a}; \cite{Ak17b}), convolving the original image by Gaussian beam with the size of the 86~GHz observation (\cite{Ha16}), and show it in figure \ref{86con}, finding it in good agreement with the observation in the scale of $\sim100r_g \approx 0.4\mathrm{mas}$ from the core.
}

\newpage

\setcounter{equation}{0}
\setcounter{figure}{0}

\textcolor{black}{
\section{Sigma cutoff}
}
\textcolor{black}{
We took no sigma cutoff in the present study, whereas the sigma cutoff, in which the region with $\sigma \equiv B^2 / 4\pi\rho_{\rm p}c^2 > \sigma_{\rm cutoff}$ is removed in transfer calculation, is often implemented in order to avoid unphysical effects arising because of low-density floors set in the MHD simulations.
To see how the results depend on the values of the sigma cutoff, we performed two additional simulations of the fiducial model (with $R_{\rm high} = 100$): one with the sigma-cutoff of $\sigma_{\rm cutoff} = 20$ and another with $\sigma_{\rm cutoff} = 1$. (The latter condition was adopted by the EHT collaboration.) Note that the mass input rate in the latter case is increased to be $\dot{M}=2.2\times 10^{-3}M_\odot{\rm yr}^{-1}$ to give the same radio flux of 0.5 Jy. The results are displayed in figure C1. \textcolor{black}{We find no large differences from that of the fiducial model shown in figure 2, and the conclusions in the present study are not altered. 
}}

\textcolor{black}{In a previous work, \cite{Ch19}, they showed different sizes of the BH-shadow among different sigma-cutoff values, since the approaching jet with high-$\sigma$ values is  brighter because of globally higher temperatures (see the central panel of their figure 3) and larger Lorentz factors. 
In our models, by contrast, the approaching jet is not so luminous to affect the size of the shadow, as seen in figure \ref{R100} and described in subsection 4.3, and the inclusion of the high-sigma region do not change the size of the shadow.}

\begin{figure*}
\begin{minipage}{0.5\hsize}
	\begin{center}
		\includegraphics[width=8cm]{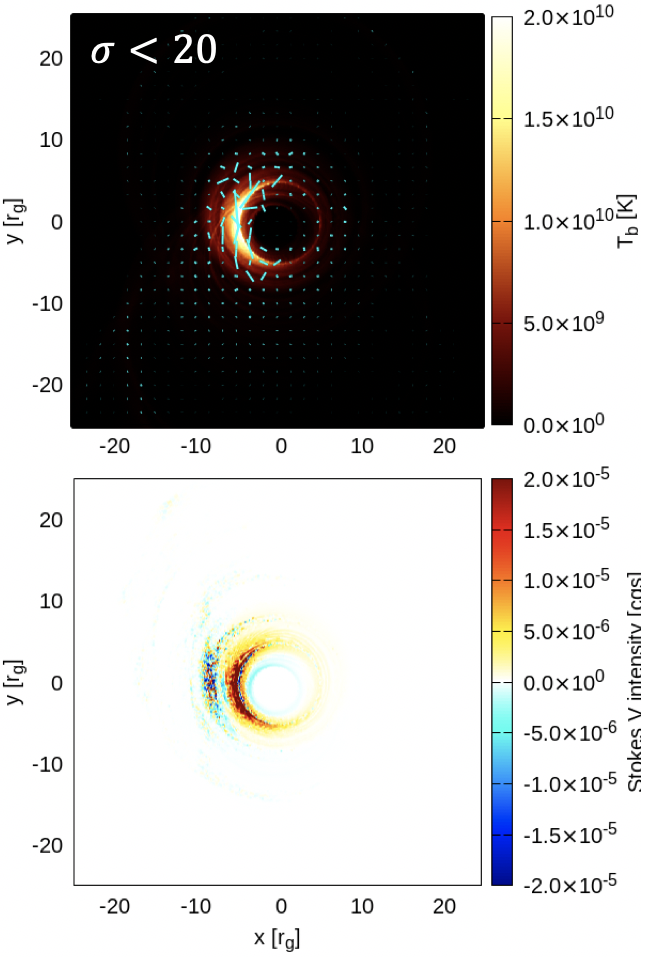}
	\end{center}
\end{minipage}
\begin{minipage}{0.5\hsize}
	\begin{center}
		\includegraphics[width=8cm]{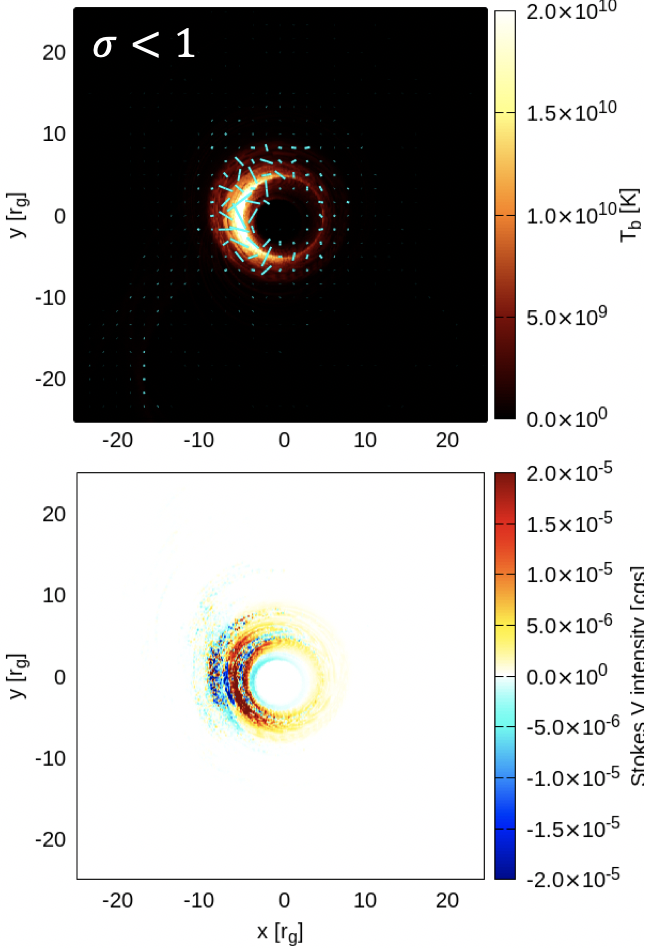}
	\end{center}	
	\end{minipage}
	\caption{\textcolor{black}{Same as figure 2 but for the case with sigma cutoff of $\sigma_{\rm cutoff} = 20$ and the case with $\sigma_{\rm cutoff} = 1$ in the left and right panels, respectively. The accretion rate is the same as that of the fiducial model in the left panel, while it is increased to be $\dot{M} = 2.2 \times 10^{-3} M_\odot {\rm yr}^{-1}$ in the right panel so as to give a total radio flux of 0.5~Jy.}}
	\label{230_sigma}
\end{figure*}


\begin{thebibliography}{99}


\bibitem[Agol (2000)]{Ag00}
	Agol, E.\ 2000, ApJL, 538, L121

\bibitem[Akiyama et al.~(2015)]{Ak15}
	Akiyama, K., Lu, R. S., Fish, V. L. et al.\ 2015, ApJ, 807, 150

\bibitem[Akiyama et al.~(2017a)]{Ak17a}
	Akiyama, K., Kuramochi, K., Ikeda, S., et al.\ 2017, ApJ, 838, 1
	
\bibitem[Akiyama et al.~(2017b)]{Ak17b}
	Akiyama, K., Ikeda, S., Pleau, M., et al.\ 2017, ApJ, 153, 159

\bibitem[Anantua et al.~(2018)]{An18}
	Anantua, R., Blandford, R., Tchekhovskoy, A.\ 2018, Galaxies, 6, 31

\bibitem[Asada \& Nakamura (2012)]{AN12}
	Asada, K., Nakamura, M.\ 2012, ApJ, 745, L28

\bibitem[Bardeen et al.~(1972)]{Bar72}
	Bardeen, J. M., Press, W. H., Teukolsky, S. A.\ 1972, ApJ, 178, 347

\bibitem[Bardeen (1973)]{Bar73}
	Bardeen, J. M.\ 1973, {\it Black holes (Les astres occlus)}, ed. C. DeWitt \& B. S. DeWitt (New York: Gordon and Breach), 215

\bibitem[Beskin (2009)]{Be09}
	Beskin, V. S.\ 2009, MHD Flows in Compact Astrophysical Objects: Accretion, Winds and Jets, Extraterrestrial Physics \& Space Sciences, Springer

\bibitem[Blandford \& K\"{o}nigl (1979)]{BK79}
	Blandford, R. D., K\"{o}nigl, A.\ 1979, ApJ, 232, 34

\bibitem[Blandford \& Znajek (1977)]{BZ77}
  Blandford, R. D., Znajek, R. L.\ 1977, MNRAS, 179, 433
  
\bibitem[Blandford \& Payne (1982)]{BP82}
  Blandford, R. D., Payne, D. G.\ 1982, MNRAS, 199, 883
  
\bibitem[Broderick \& Loeb (2009)]{BL09}
  Broderick, A. E., Loeb, A.\ 2009, ApJ, 697, 1164

\bibitem[Carter (1968)]{Ca68}
	Carter, B.\ 1968, Physical Review, 174, 1559
	
\bibitem[Chael et al.~(2016)]{Ch16}
	Chael, A. A., Johnson, M. D., Narayan, R. et al.\ 2016, ApJ, 829, 11

\bibitem[Chael et al.~(2019)]{Ch19}
	Chael, A., Narayan, M., Johnson, M. D.\ 2019, MNRAS, 486, 2873

\bibitem[Chandrasekhar (1983)]{Ch83}
	Chandrasekhar, S. \ 1983, The Mathematical Theory of Black Holes, Clarrendon Press/Oxford Univ. Press, Oxford/New York

\bibitem[Dexter et al.~(2012)]{De12}
	Dexter, J., McKinney, J. C., Agol, E.\ 2012, MNRAS, 421, 1517

\bibitem[Dexter (2016)]{De16}
  Dexter, J.\ 2016, MNRAS, 462, 115
  
\bibitem[Doeleman et al.~(2012)]{Do12}
	Doeleman, S. S., Fish, V. L., Schenck, D. E. et al.\ 2012, Science, 338, 355  
  
\bibitem[EHT Collaboration et al.~(2019a)]{EHT19a}
  Event Horizon Telescope Collaboration et al.\ 2019a, ApJL, 875, L1 (Paper \Rnum{1})

\bibitem[EHT Collaboration et al.~(2019d)]{EHT19d}
  Event Horizon Telescope Collaboration et al.\ 2019d, ApJL, 875, L4 (Paper \Rnum{4})

\bibitem[EHT Collaboration et al.~(2019e)]{EHT19e}
  Event Horizon Telescope Collaboration et al.\ 2019e, ApJL, 875, L5 (Paper \Rnum{5})

\bibitem[Falcke et al.~(2000)]{Fa00}
	Falcke, H., Melia, F., Agol, E.\ 2000, ApJ, 528, L13

\bibitem[Fukue \& Yokoyama (1988)]{FY88}
	Fukue, J., Yokoyama, T.\ 1988, PASJ, 40, 15

\bibitem[Gammie et al.~(2003)]{HARM}
	Gammie, C. F., McKinney, J. C., T\'{o}th, G.\ 2003, ApJ, 589, 444

\bibitem[Gebhardt et al.~ (2011)]{Ge11}
	Gebhardt, K.,  Adams, J.,  Richstone, D. et al.\ 2011, ApJ, 729, 119

\bibitem[Hada et al.~(2013)]{Ha13}
	Hada, K., Kino, M., Doi, A. et al.\ 2013, ApJ, 775, 70

\bibitem[Hada et al.~(2016)]{Ha16}
	Hada, K., Kino, M., Doi, A. et al.\ 2016, ApJ, 817, 131

\bibitem[Hilbert (1916)]{Hi16}
	Hilbert, D.\ 1916, Die Grundlagen der Physik. (Zweite Mitteilung), Nachrichten von der K\"oniglichen Gesellschaft der Wissenschaften zu G\"ottingen. Math. phys. Klasse. 1917 (Berlin: Weidmannsche Buchhandlung), 53
	
\bibitem[Kato et al.~(2008)]{KFM08}
	Kato, F., Fukue, J., Mineshige, S.\ 2008, ``Black-Hole Accretion Disks 2nd edition: Towards a New Paradigm'', Kyoto University Press

\bibitem[Kim et al.~(2016)]{Ki16}
	Kim, J. Y., Lu, R. S., Krichbaum, T. et al.\ 2016, Galaxies, 4, 39
	
\bibitem[Kino et al.~(2015)]{Ki15}
	Kino, M., Takahara, F., Hada, K. et al.\ 2015, ApJ, 803, 30
	
\bibitem[Krolik et al.~(2005)]{Kro05}
	Krolik, J. H., Hawley, J. F., Hirose, S.\ 2005, ApJ, 622, 1008
	
\bibitem[Kulkarni et al.~(2011)]{Ku11}
	Kulkarni, A. K., Penna, R. F., Shcherbakov, R. V. et al.~\ 2011, MNRAS, 414, 1183

\bibitem[Kuo et al.~(2014)]{Kuo14}
  Kuo, C. Y., Asada, K., Rao, R. et al.\ 2014, ApJ, 783, L33

\bibitem[Luminet (1979)]{Lu79}
	Luminet, J. P.\ 1979, Astronomy \& Astrophysics, 75, 228

\bibitem[Mahadevan et al.~(1996)]{Ma96}
	Mahadevan, R., Narayan, R., Yi, I.\ 1996, ApJ, 465, 327

\bibitem[Marrone et al.~(2006)]{Ma06}
	Marrone, D. P., Moran, J. M., Zhao, J. H., Rao, R.\ 2006, ApJ, 640, 308

\bibitem[Mei et al.~(2007)]{Mei07}
	Mei, S., Blakeslee, J. P., C\^ot\'e, P.,  Tonry, J. L. et al.\  2007, ApJ, 655, 144

\bibitem[Mo\'{s}cibrodzka et al.~(2016)]{Mo16}
	Mo\'{s}cibrodzka, M., Falcke, H., Shiokawa, H.\ 2016, Astronomy \& Astrophysics, 586, A38

\bibitem[Mo\'{s}cibrodzka et al.~(2017)]{Mo17}
  Mo\'{s}cibrodzka, M., Dexter, J., Davelaar, J., Falcke, H.\ 2017, MNRAS, 468, 2214
  
\bibitem[Mo\'{s}cibrodzka et al.~(2018)]{Mo18}
  Mo\'{s}cibrodzka, M., Gammie, C. F.\ 2018, MNRAS, 475, 43
  
\bibitem[Moriyama \& Mineshige (2015)]{MM15}
	Moriyama, K., Mineshige, S.\ 2015, PASJ, 67, 106

\bibitem[Nagar et al.~(2000)]{Nag00}
	Nagar, N. M., Falcke, H., Wilson, A. S., Ho, L. C.\ 2000, ApJ, 542, 186

\bibitem[Nakamura \& Asada (2013)]{NA13}
	Nakamura, M., Asada, K.\ 2013, ApJ, 775, 118

\bibitem[Nakamura et al.~(2018)]{Na18}
  Nakamura, M., Asada, K., Hada, K. et al.\ 2018, ApJ, 868, 146

\bibitem[Shcherbakov (2008)]{Sh08}
  Shcherbakov R. V.\ 2008, ApJ, 688, 695
  
\bibitem[Shcherbakov \& Huang (2011)]{SH11}
	Shcherbakov R. V., Huang, L.\ 2011, MNRAS, 412, 1052
  
\bibitem[Shcherbakov et al.~(2012)]{Sh12}
  Shcherbakov, R. V., Penna, R. F., McKinney, J. C.\ 2012, ApJ, 755, 133

\bibitem[Takahashi (2004)]{Ta04}
	Takahashi, R.\ 2004, ApJ, 611, 996
	
\bibitem[Tchekhovskoy et al.~(2011)]{Tch11}
	Tchekhovskoy, A., Narayan, R., McKinney, J. C.\ 2011, MNRAS, 418, L79

\bibitem[von Laue (1921)]{von21}
	von Laue, M.\ 1921, Relativit\"atstheorie, Vol. 1 (Braunschweig: FriedrichVieweg \& Sohn) Das Relativit\"atsprinzip der Lorentztransformation - Vol. 2.Die allgemeine Relativit\"atstheorie und Einsteins Lehre von der Schwerkraft

\bibitem[Walker et al.~(2018)]{Wa18}
  Walker, R. C., Hardee, P. E., Davies, F. B. et al.\ 2018, ApJ, 855, 128
     
\end{thebibliography}
\end{document}